\documentclass{aa}  
\usepackage{lscape}
\usepackage{hyperref}
\hypersetup{bookmarks=true, colorlinks=true, citecolor=blue, linkcolor=blue}
\usepackage{graphicx}
\usepackage[table, svgnames, dvipsnames]{xcolor}
\usepackage{txfonts}
\usepackage{color}

\def\msol{\ensuremath{M_\odot}}

\newcommand\ST[1]{{\color{black}#1}}
\def\RM#1{{\color{black} #1}}
\begin{document}

   \title{Chemical Evolution of R-process Elements in Stars (CERES)}

   \subtitle{II. The impact of stellar evolution and rotation on light and heavy elements}

   \author{Raphaela Fernandes de Melo\inst{1} \and
          Linda Lombardo\inst{1}\and
          Arthur Alencastro Puls\inst{1}\and
          Donatella Romano\inst{2}\and
          Camilla Juul Hansen\inst{1}\and
          Sophie Tsiatsiou\inst{3}\and        
          Georges Meynet\inst{3}}
          
     %Department of Astronomy, University of Geneva, Chemin Pegasi 51, 1290 Versoix, Switzerland for Georges Meynet, Sophie Tsiatsiou

    \institute{Goethe University Frankfurt, Institute for Applied Physics, Max-von-Laue-Str. 12,
60438 Frankfurt am Main, Germany\\
                \email{FernandesdeMelo@physik.uni-frankfurt.de}
                \and
                INAF, Osservatorio di Astrofisica e Scienza dello Spazio, Via Gobetti 93/3, I-40129 Bologna, Italy
                \and 
                Department of Astronomy, University of Geneva, Chemin Pegasi 51, 1290 Versoix, Switzerland
                }

%   \institute{Institute for Astronomy (IfA), University of Vienna,
%              T\"urkenschanzstrasse 17, A-1180 Vienna\\
%              \email{wuchterl@amok.ast.univie.ac.at}
%         \and
%             University of Alexandria, Department of Geography, ...\\
%             \email{c.ptolemy@hipparch.uheaven.space}
%             \thanks{The university of heaven temporarily does not
%                     accept e-mails}
%             }
%
%   \date{Received September 15, 1996; accepted March 16, 1997}

% \abstract{}{}{}{}{} 
% 5 {} token are mandatory
 
  \abstract
   {Carbon, nitrogen, and oxygen are the most abundant elements throughout the universe, after hydrogen and helium. Studying these elements in low-metallicity stars can provide crucial information on the chemical composition in the early Galaxy and possible internal mixing processes that can alter the surface composition of the stars.}
   {This work aims to investigate the chemical abundance patterns for CNO elements and Li in a homogeneously analyzed sample of 52 metal-poor halo giant stars. From these results, we have been able to determine whether internal mixing processes have taken place in these stars.}
   {We used high-resolution spectra with a high signal-to-noise ratio (S/N) to carry out a spectral synthesis to derive detailed C, N, O, and Li abundances for a sample of stars with metallicities in the range  of$-$3.58 $\leq$ [Fe/H] $\leq$ $-$1.79\,dex. Our study was based on the assumption of one-dimensional (1D) local thermodynamic equilibrium (LTE) atmospheres.}
   {Based on carbon and nitrogen abundances, we investigated the deep mixing taking place within stars along the red giant branch (RGB). The individual abundances of carbon decrease towards the upper RGB while nitrogen shows an increasing trend, indicating that carbon has been converted into nitrogen. No signatures of ON-cycle processed material were  found for the stars in our sample. 
   We computed a set of galactic chemical evolution (GCE) models,  implementing different sets of massive star yields, both with and without including the effects of stellar rotation on nucleosynthesis. We confirm that stellar rotation is necessary to explain the highest $[$N$/$Fe$]$ and $[$N$/$O$]$ ratios observed in unmixed halo stars. The predicted level of N enhancement varies sensibly in dependence of the specific set of yields that are adopted. For stars with stellar parameters similar to those of our sample,  heavy elements such as Sr, Y, and Zr appear to have unchanged abundances despite the stellar evolution mixing processes.}
   {The unmixed RGB stars provide very useful constraints on chemical evolution models of the Galaxy.
   As they are more luminous than unevolved (main sequence and turnoff) stars, they also allow for stars to be probed at greater distances. The stellar CN-cycle clearly changes the atmospheric abundances of the lighter elements, but no changes were detected with respect to the heavy elements.}


   \maketitle
%
%-------------------------------------------------------------------

\section{Introduction}

Carbon, nitrogen, and oxygen are among the most abundant elements throughout the Universe.
A well-established fact in nuclear astrophysics is that carbon, nitrogen, and oxygen are synthesized through hydrostatic nuclear fusion occurring in different stages of stellar evolution or by explosive burning taking place in the end of the evolution of massive stars \citep{1957RvMP...29..547B, 2013ARA&A..51..457N, 2020ApJ...900..179K, 2022A&ARv..30....7R, 2023A&ARv..31....1A}. The nucleosynthetic processes in the stellar interior that lead to the production of these elements are also \RM{well known} and depend on the mass, metallicity, and rotation of the star, with the role of binary interactions yet to be fully investigated \citep[e.g., ][]{2021ApJ...923..214F}.

The stable isotopes of carbon and oxygen are primarily synthesized by hydrostatic He burning through the triple-$\alpha$ reaction \citep{1954ApJS....1..121H, 2002RvMP...74.1015W}. Although carbon is produced in stars of all masses, asymptotic giant branch (AGB) star models estimate a contribution of roughly one third of the $^{12}$C found in the Galaxy \citep{2014PASA...31...30K}
This is in agreement with \citet{2020A&A...639A..37R} finding that 60-70\% of C being formed by massive stars\footnote{The exact fraction depends on the yields adopted but also on the host galaxy.}.
On the other hand, the main contribution to the oxygen abundance comes from massive stars via core-collapse supernovae (CCSNe) \citep{2016MNRAS.458.3466V}. Oxygen is the most abundant metal in stars.  Nitrogen is mainly a secondary element and mostly produced by low- and intermediate-mass stars through the CN-cycle at the expense of the carbon and oxygen initially present in the star. If nuclear burning at the base of the convective envelope is efficient, nitrogen can also be produced as a primary element from the original hydrogen and helium during the third dredge-up (TDU) in AGB stars \citep{1981A&A....94..175R}. 

Stellar models have shown that at \RM{low metallicities}, rotation  also has a significant impact on the observed abundance of the CNO elements on the surface of the stars. This is due to the fact that the mixing of chemical elements becomes more efficient for a given initial velocity, implying the production of large quantities of primary nitrogen in the H-burning shell \citep[see e.g.,][]{2001A&A...373..555M,2002A&A...390..561M, 2006A&A...447..623M,2006A&A...449L..27C, 2007A&A...461..571H, 2018ApJS..237...13L, 2024arXiv240416512T}.

Elemental abundances for C, N, and O have been derived for stars in the disk and metal-poor halo \citep[e.g.,][]{2004A&A...421..649I, 2004A&A...416.1117C, 2004A&A...414..931A, 2005A&A...430..655S,2019A&A...622L...4A}. With regard to the metal-poor halo stars, \citeauthor{2019A&A...622L...4A} derived homogeneous carbon and oxygen abundances for a sample of 39 metal-poor turn-off stars with $-$3.0 \RM{$<$ [Fe/H] $< -$1.0~dex}, taking into account three-dimensional (3D) non-local thermodynamic equilibrium (non-LTE) effects, and found flat [C/Fe] ratios relatively to [Fe/H], while [O/Fe] increases linearly with decreasing [Fe/H].

Chemical abundance analyses resting on high-resolution spectroscopy of stars in the Local Group produce reliable CNO element abundances and these are crucial in several astrophysical fields, including stellar astrophysics as they can provide constrains on nucleosynthesis theory and stellar evolution \citep{2018A&ARv..26....6N}, playing a key role to understand galactic chemical evolution and stellar populations. Additionally, these elements can also be used to investigate internal mixing processes taking place in the stars during the giant branch phase, which are responsible for transporting material from the stellar interior to the outer layers and vice versa, shaping the final surface elemental abundances observed in the stars \citep{Korn2006Natur, Korn2007}.

However, the explanation for the CNO production in stars is not yet satisfactory and a self-consistent explanation for the CNO abundances measured for diverse populations of stars in the Galaxy is required from galactic chemical evolution models.
This paper targets CNO in metal-poor halo stars to assess the impact of internal mixing processes and the chemical evolution of the halo.
The paper is \RM{organized} as follows. The stellar sample and data reduction are described in Sect.~\ref{sec:sample}, the atmospheric parameters in Sect.~\ref{sec:parameters}, and the abundances in Sect.~\ref{sec:abundances}. Our results and discussion appear in Sects.~\ref{sec:results} and \ref{sec:discussion}, respectively. Finally, our conclusions are given in Sect.~\ref{sec:conclusion}.

\section{Sample and data reduction \label{sec:sample}}

The sample consists of 52 giant stars of \RM{low metallicity} ([Fe$/$H] $< -1.5$~dex). To avoid contamination from a companion, only stars with no evidence of binarity were considered; thus, stars with clear radial velocity variation were not considered further, as detailed in \citet[][hereafter Paper\,I]{2022A&A...665A..10L}. The selection criteria also reject stars classified as carbon-enhanced metal-poor (CEMP) stars given the difficulty of measuring accurately heavy element abundances due to the strong CH and CN molecular bands. 

Observations %of the target stars 
were carried out during two runs (November 2019 and March 2020) using the high-resolution Ultraviolet and Visual Echelle Spectrograph \citep[UVES;][]{2000SPIE.4008..534D} of the ESO Very Large Telescope (VLT) in Cerro Paranal, Chile. The target stars were observed  with an 1\arcsec~slit, 1x1 binning and with the standard UVES setup DIC. Central wavelengths 390 and 564 nm were used for the blue and red arms, respectively, resulting in an average resolving power of R $\sim 49,800$ in the blue arm and R $\sim 47,500$ in the red arm. Details of the observations for the individual stars are provided in the appendix of Paper\,I.

\section{Atmospheric parameters \label{sec:parameters}}

As described in Paper I, the atmospheric parameters were estimated based on $G_{BP} - G_{RP}$ photometry and parallaxes from \textit{Gaia} Early Data Release 3 \citep{2016A&A...595A...1G, 2021A&A...649A...1G}. For effective temperature (\RM{$T_\mathrm{eff}$}) and surface gravities (log$\,g$), the iterative procedure described in \cite{2021A&A...645A..64K} was employed. \RM{Microturbulent} velocities (\RM{$v_\mathrm{turb}$}) were obtained based on the calibration from \citet{Mashonkina2017A&A...604A.129M}. Metallicities were derived from \ion{Fe}{i} lines using the code MyGIsFOS \citep{2014A&A...564A.109S}. The final stellar parameters are in Table~\ref{parameters}.
The adopted uncertainties for \RM{$T_\mathrm{eff}$}, log$\,g$, \RM{$v_\mathrm{turb}$}, and [Fe/H] are respectively 100$\,$K, 0.04$\,$dex, 0.5$\,$km$\,$s$^{-1}$, and 0.13$\,$dex, as stated in Paper$\,$I.

\section{Abundances \label{sec:abundances}}

The abundance analysis for all the stars is carried out using the driver ``synth'' of the LTE stellar line synthesis program \texttt{MOOG}\footnote{\url{https://www.as.utexas.edu/~chris/moog.html}} (\citealp[]{1973PhDT.......180S}, version 2019) combined with ATLAS12 model atmospheres \citep{2005MSAIS...8...14K}. When fitting the synthetic spectrum to the observed one, the abundances from Paper I were also taken into account. 
Species studied in Paper I that were not detected in a particular star were treated as absent ([X/Fe] = -9.99\RM{~dex}). 
The adopted solar abundances for C, N, and O were taken from \citet{2009LanB...4B..712L}, with their values listed in Table~\ref{solar}. The values for oscillator strengths and lower excitation energies were taken from line lists generated with \texttt{Linemake}\footnote{\url{https://github.com/vmplacco/linemake}} \citep[][]{2021RNAAS...5...92P}. When considering blending, these line lists take into account all molecular species available in \texttt{Linemake}, except for TiO containing isotopes other than \element[ ][48]{Ti}.

\begin{table}
    \caption{Adopted Solar abundances in this study. 
    }
\begin{tabular}{ccc}
\hline\hline
Element & A(X) & ref.\\\hline 
Li & 1.10 &  \cite{2009LanB...4B..712L} \\
C  & 8.39 &  \cite{2009LanB...4B..712L} \\
N  & 7.86 &  \cite{2009LanB...4B..712L} \\
O  & 8.73 &  \cite{2009LanB...4B..712L} \\
\hline 
Comparison elements \\
\hline
Mg & 7.54 &  \cite{2009LanB...4B..712L} \\
Sr & 2.92 &  \cite{2009LanB...4B..712L} \\
Y  & 2.21 &  \cite{2009LanB...4B..712L} \\
Zr & 2.62 &  \cite{2011AN....332..128C} \\
\hline
\end{tabular}%
\tablefoot{The comparison elements refer to the elements used in this paper and obtained in Paper\,I.
}
\label{solar}
\end{table}

The sensitivities of C, N, and O abundances were determined in the following way. Eight models were calculated for a representative star, \RM{CES\,0031$-$1647}, changing each of the four atmospheric parameters individually by their uncertainties. \RM{CES\,0031$-$1647} was chosen because (among all the targets with measurements for C, N and O) its stellar parameters have the shortest Euclidean distance to the median point of the log(\RM{$T_\mathrm{eff}$})-log$\,g$-[Fe/H] space of our sample. The C, N, and O sensitivities were calculated by fitting the spectra synthesized with these eight models (listed in Table~\ref{tab:sensitivities}). The same procedure has been adopted for computing Li sensitivities using the star \RM{CES\,0045$-$0932} instead of \RM{CES\,0031$-$1647} because in the latter, Li was not measurable. The uncertainties for the elemental abundance
ratios with iron were computed according to the procedure explained in the
appendix of \cite{1995AJ....109.2757M}.

\begin{table}
    \centering
    \caption{Computed sensitivities for CNO and Li abundances  with respect to the uncertainties in the stellar parameters. }
\begin{tabular}{lrrrr}
\hline\hline
 & $\Delta$A(C) & $\Delta$A(N) & $\Delta$A(O) & $\Delta$A(Li)\\
 & dex & dex & dex & dex \\
\hline
\RM{$T_\mathrm{eff}$} $+$ 100~K               & $+$0.23 & $+$0.22 & $+$0.07 & $+$0.11 \\
\RM{$T_\mathrm{eff}$} $-$ 100~K               & $-$0.23 & $-$0.22 & $-$0.07 & $-$0.10 \\
log$\,g$ $+$ 0.04~dex                         & $-$0.02 & $+$0.02 & $+$0.01 & <0.01 \\
log$\,g$ $-$ 0.04~dex                         & $+$0.01 & $-$0.03 & $-$0.01 & $+$0.01 \\
\RM{$v_\mathrm{turb}$} $+$ 0.5~km$\,$s$^{-1}$ & <0.01   & $-$0.05 & <0.01   & <0.01 \\
\RM{$v_\mathrm{turb}$} $-$ 0.5~km$\,$s$^{-1}$ & $-$0.01 & $+$0.05 & $+$0.01 & $+$0.01 \\
$[$Fe/H$]$ $+$ 0.13~dex                       & $-$0.01 & <0.01   & $+$0.01 & $+$0.01 \\
$[$Fe/H$]$ $-$ 0.13~dex                       & $+$0.02 & <0.01   & $-$0.01 & <0.01 \\
\hline
\end{tabular}
\tablefoot{Sensitivities of CNO abundances in star \RM{CES\,0031$-$1647}. For Li, the sensitivities are computed for the star \RM{CES\,0045$-$0932}.

}
\label{tab:sensitivities}
\end{table}

\begin{figure*}
\sidecaption
        \includegraphics[width=12cm]{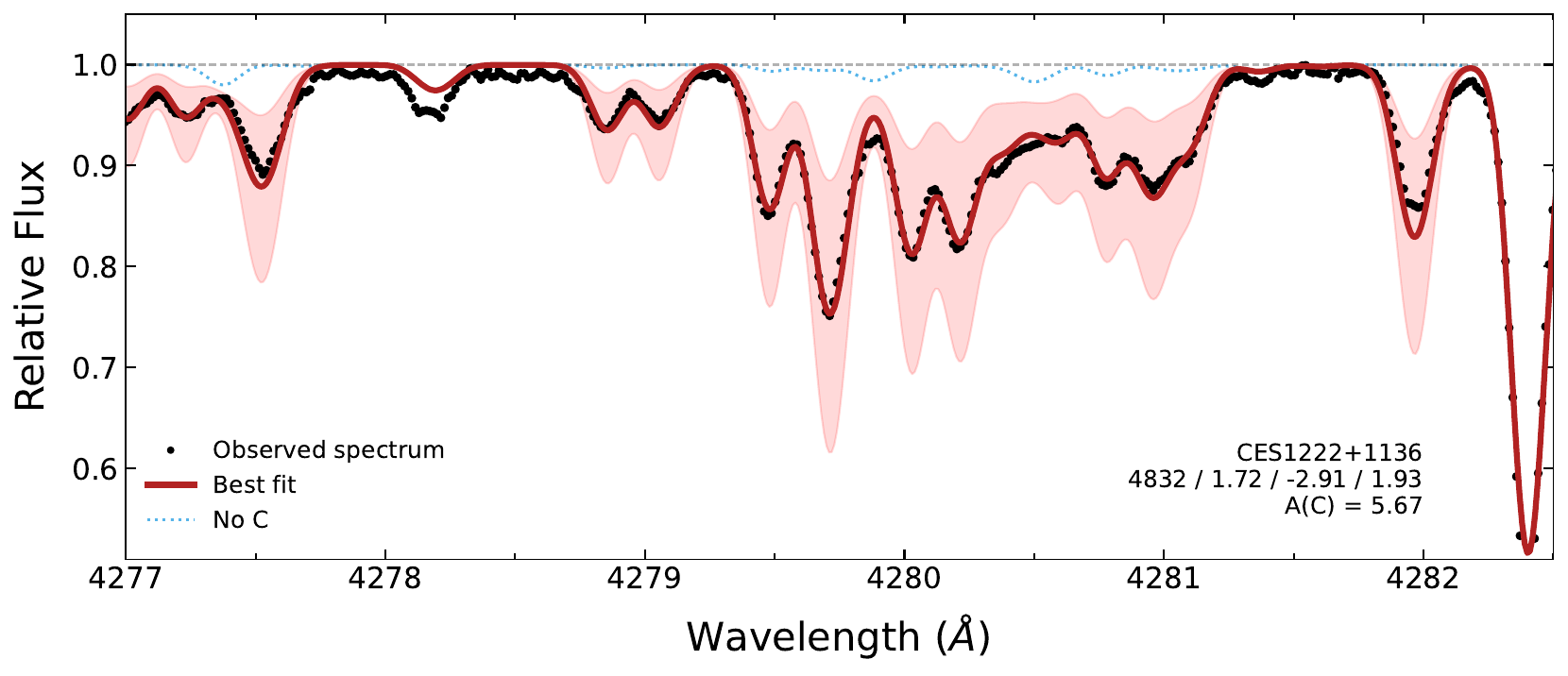}
    \caption{
    Portion of the spectrum of star CES\,1222$+$1136, showing the G band on the region 4277-4282\,\AA. 
    The solid curves show the best fit with A(C) = 5.67 %\textbf{Linda: this values is different from the one in the Figure} 
    (thicker red line), and the red shaded areas  display A(C) $\pm$ 0.30\,dex.
    The dotted %{\bf GM dotted instead of dashed?} 
    blue line represents the synthesis if C is fully removed.
    The atmospheric parameters \RM{$T_\mathrm{eff}$} [K], \RM{$\mathrm{log}\,g$}, [Fe/H], and \RM{$v_\mathrm{turb}$} [km s$^{-1}$] are shown in the lower right corner. %{\bf GM the label on the right lower corner in front of blue dotted line should be No C.}
    }
    \label{fig:ch_4277}
\end{figure*}

\begin{figure*}
\sidecaption
        \includegraphics[width=12cm]{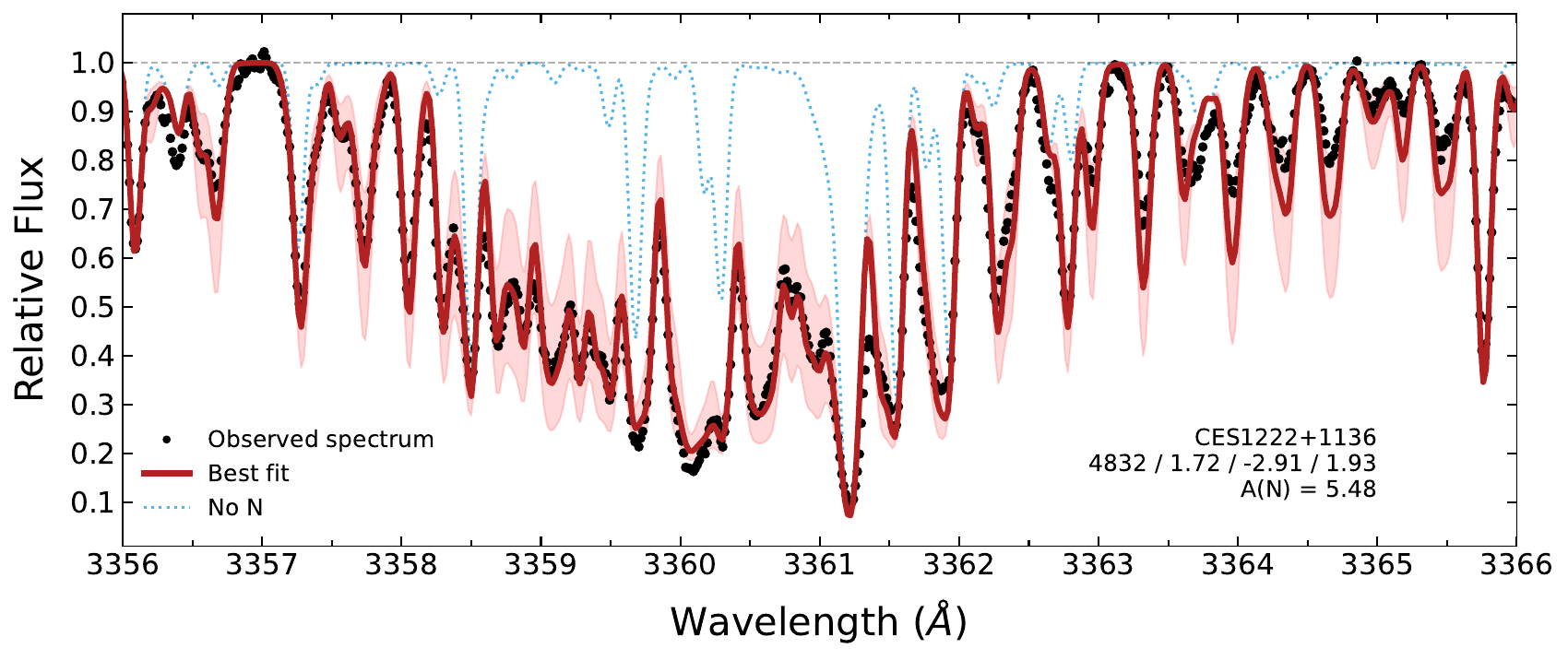}
    \caption{
    Portion of the spectrum of star CES\,1222$+$1136, showing the NH band at 3360\,\AA.
    The solid curves show the best fit with A(N) = 5.48 %\textbf{Linda: same as Fig.1, the abundance does not correspond to the value in figure}
    (thicker red line), and the red shaded areas  display A(N) $\pm$ 0.30\,dex.
    The dashed blue line represents the synthesis if N is fully removed.
    The atmospheric parameters \RM{$T_\mathrm{eff}$} [K], \RM{$\mathrm{log}\,g$}, [Fe/H], and \RM{$v_\mathrm{turb}$} [km s$^{-1}$] are shown in the lower right corner.
    }
    \label{fig:nh_3360}
\end{figure*}

\subsection{Lithium}
For nine stars in the sample, we were able to measure the lithium abundances from the Li resonance line at 6707\,\AA\ by fitting the observed line profile with synthetic spectra computed with MOOG. The adopted oscillator strength value as well as the full isotopic and hyperfine substructure of the line were taken from the Kurucz database\footnote{\url{http://kurucz.harvard.edu/linelists.html}}. 
The derived 1D LTE Li abundances are listed in Table~\ref{tab:li_abundances}.
In the same table, we also provide Li abundances obtained by applying 3D non-local thermodynamic equilibrium (NLTE) corrections provided by \citet{2021MNRAS.500.2159W}. The 3D NLTE corrections listed in Table~\ref{tab:li_abundances} are small, varying from $-$0.01 to $+$0.04\,dex; \RM{hence, they are much lower than the typical error associated to our Li abundance determinations.}

\begin{table}
    \centering
    \caption{Lithium abundances and 3D NLTE corrections obtained for the stars in this work.}
\begin{tabular}{lcc}
\hline\hline
  Star         & A(Li) & A(Li)$_{3D\,NLTE}$ \\
  \hline
  CES\,0045$-$0932 & 1.01   & 1.03         \\
  CES\,0338$-$2402 & 1.03   & 1.07         \\
  CES\,1221$-$0328 & 0.94   & 0.98         \\
  CES\,1427$-$2214 & 0.94   & 0.93         \\
  CES\,1436$-$2906 & 1.05   & 1.06         \\
  CES\,1543$+$0201 & 1.04   & 1.07         \\
  CES\,2231$-$3238 & 1.04   & 1.07         \\
  CES\,2232$-$4138 & 1.13   & 1.15         \\
  CES\,2330$-$5626 & 0.96   & 0.98         \\\hline
\end{tabular}
\label{tab:li_abundances}
\end{table}

\begin{figure}[h!]
  \resizebox{\hsize}{!}{\includegraphics{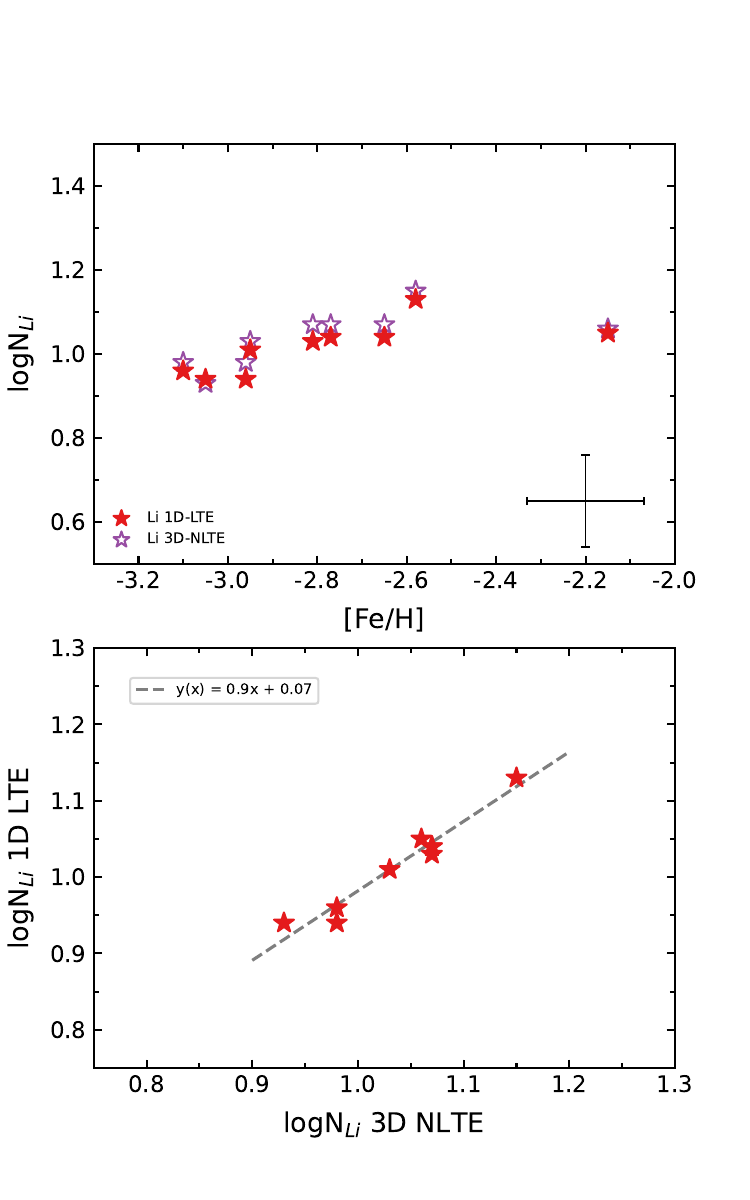}}
  \caption{ Lithium abundances obtained with our 1D LTE analysis (red symbols) and corrected for 3D NLTE effects (violet open symbols) as a function of [Fe/H] for the stars in our sample (top). A representative error bar is plotted in the lower-right corner of the figure. In the lower panel, we compare both 1D LTE and 3D NLTE lithium abundances. The stars CES\,1543$+$0201 and CES\,2231$-$3238 overlap around log\,N$_{Li}$ = 1.04 and log\,N$_{Li}$ 3D NLTE = 1.07.}
  \label{fig:li_fe}
\end{figure}

\subsection{Carbon}
The carbon abundances were derived by fitting the synthetic spectrum to the CH lines of the band $A^2\Delta - X^2$ (the G band). We \RM{analyzed} small regions between 4277-4330\,\AA\,, focusing mainly on the regions 4277-4282\,\AA\ and 4305-4315\,\AA\,. 
This region was used because it is sensitive to C and there are only a few atomic lines in this range. 
The adopted excitation potential, $\log gf$, and dissociation energies for CH lines were taken from \citet{Masseron2014A&A...571A..47M}. 

\RM{It is worth mentioning that three-dimensional (3D) hydrodynamical corrections for red giant stars may affect the strength of spectral lines and favor a higher concentration of CH, NH, and OH molecules. At [Fe/H] =$-3$, abundances for C, N, and O derived from these molecules from 3D (LTE) computations are $\sim$0.5\,dex to $\sim$1.0\,dex lower than 1D (LTE) abundances \citep{2007A&A...469..687C}.} Recent non-LTE studies \citep[e.g.,][]{2023A&A...670A..25P} have shown that the corrections to CH increase with decreasing metallicity, reaching +0.21 at [Fe/H] = $-4$\RM{~dex} in red giants. A star with T=4500~K, log$\,g$ = 2.0~\RM{dex} and metallicities from \RM{0 down to $-4$~dex} will have corrections from \RM{$\sim 0.12$ to $\sim 0.21$~dex}, respectively. This example is representative of our sample stars and we therefore \RM{should} expect non-LTE corrections of the order of $\leq 0.2$\,dex for our LTE CH abundances. As we do not have any C-enhanced stars in the sample, we expect the corrections to remain low and positive. 

\subsection{Nitrogen}

Nitrogen abundances were calculated using the $A^3\Pi - X^3\Sigma$ NH band at 3360\,\AA\,. We fit the 3357$-$3363\,\AA\ interval, taking into account the abundances from Paper I. 
The adopted dissociation energy for the NH molecule is 3.47\,eV \citep[][]{Huber1979}, which is the same used by \citet[][]{2005A&A...430..655S}.

Figure~\ref{fig:nh_3360} presents an example of spectral synthesis in the N-poor star CES\,1222$-$1136. The best fit shows agreement in most of the evaluated wavelength range, with notable exceptions in a few lines where the synthetic spectrum with N-enhancement of 0.30\,dex has a better agreement; an additional exception is seen at 3360\,\AA\,, where the best fit slightly overpredicts the observed spectrum. Interestingly, the spectral fit of the NH band shown in Fig.~1 from \citet[][]{2005A&A...430..655S} for star BD$-$18\,5550 has similar deviations in the same wavelength range. While our evaluation interval between 3357 and 3363\,\AA\ aims to average out these deviations in spectral fitting, the under or overfitting, similar to that shown in \citet[][]{2005A&A...430..655S}, suggests the need to revise the atomic and molecular data of NH and the other species contributing to the blends.

Removing N from the synthetic spectrum in the 3360\,\AA\ band suggests that among the least blend-affected lines used to evaluate N (at least in metal-poor stars) are those located at 3358.05\,\AA, 3359.07\,\AA, and 3359.28\,\AA\ ( as  seen in Fig.~\ref{fig:nh_3360}). They are not affected by species with the strongest features (after N) in that wavelength region, such as Cr, which lack measurements in a few of our stars. Hence, assuming that their atomic data are accurate, they are reliable to check the validity of the spectral fit, in particular in these stars lacking measurements of elements blending with the N band. % (e.g., CES\,2103$-$6505).
We tested the cases when Ca, Sc, Cr, or Zr were entirely removed from the line list in the most metal-rich star under study (CES\,0424$-$1501).  The impact in the N abundance is lower than 0.05~dex.

\subsection{Oxygen} \label{oxygen}
Oxygen abundances were derived from the \RM{[\ion{O}{i}]} lines at 6300.304\,\AA\ and 6363.776\,\AA\,. The adopted $gf$-values for these transitions are $\log gf = -9.72$ and $\log gf = -10.19$, respectively \citep{2000MNRAS.312..813S}.
These lines are generally considered the most reliable for deriving O abundances, since they are unaffected by NLTE effects. However, it has been shown that these lines are sensitive to 3D effects, which tend to become larger \RM{toward} lower metallicities  \citep[see e.g.,][]{Nissen2002A&A...390..235N,Amarsi2016MNRAS.455.3735A}.
The forbidden oxygen lines in metal-poor stars are weak, so we could determine O abundances only in a subsample of our stars (19 of 52). The measurement of the \RM{[\ion{O}{i}]} line at 6300\,\AA\ is also complicated by the presence of telluric features in the same spectral region, which can contaminate the oxygen line profile.
The obtained abundances are listed in Table~\ref{tab:abundances}.

\RM{As there are no recent NLTE corrections for the NH lines and the available corrections for CH do not cover the atmospheric parameters range of most of our stars, in the remainder of this paper, we consider the 1D LTE results we previously derived.}

\section{Abundance results \label{sec:results}}

The nine stars in our sample for which the Li line is measurable show very similar lithium abundances, with an average A(Li) = 1.04$\,\pm\,$0.1~dex and low dispersion of $\sigma$ = 0.06~dex. Our results for lithium at low metallicity are in good agreement with the measurements of \cite{2022A&A...661A.153M}, where an average A(Li) = 1.06$\,\pm\,$0.01 dex was found for 47 carbon-normal stars on the lower red giant branch (LRGB) in the metallicity range of $-$3.80 $\leq\,$[Fe/H]$\,\leq-$1.3~dex.  
It should also be highlighted that seven out of the nine stars for which we derived lithium abundances were also \RM{analyzed} by \cite{2022A&A...661A.153M}. Their results are in agreement  with those found in this study within the errors.

The upper panel of Fig.~\ref{fig:li_fe} shows that A(Li) increases with increasing metallicity and the slope for the plot is equal to 0.14. If the star \RM{CES\,1436$-$2906}, with \RM{[Fe/H] = $-$2.15~dex}, is excluded, the mean average for A(Li) does not change significantly while the slope increases to 0.31. The presence of this linear trend for lithium is possibly attributed to the small subsample of stars with Li features in our sample and could be erased by increasing the sample size. Moreover, the changes in the abundance of lithium in the considered metallicity domain do fit inside  the error bar indicated in the lower right corner of the figure; therefore, the small linear change with metallicity is also compatible with a constant Li abundance.   
Adopting the 3D NLTE corrections, we obtained Li abundances slightly higher in comparison with the non-corrected abundances (lower panel in Fig.~\ref{fig:li_fe}). In fact, the largest correction applied is 0.04\RM{~dex} in A(Li), and the linear fit that correlates the uncorrected Li abundances with the corrected ones has a slope of $\sim\,$0.91. Therefore, this difference (3D, NLTE correction) does not change the slope for A(Li) versus [Fe/H] and is not significant for our results.

\begin{figure*}
\centering
   \includegraphics[width=17cm]{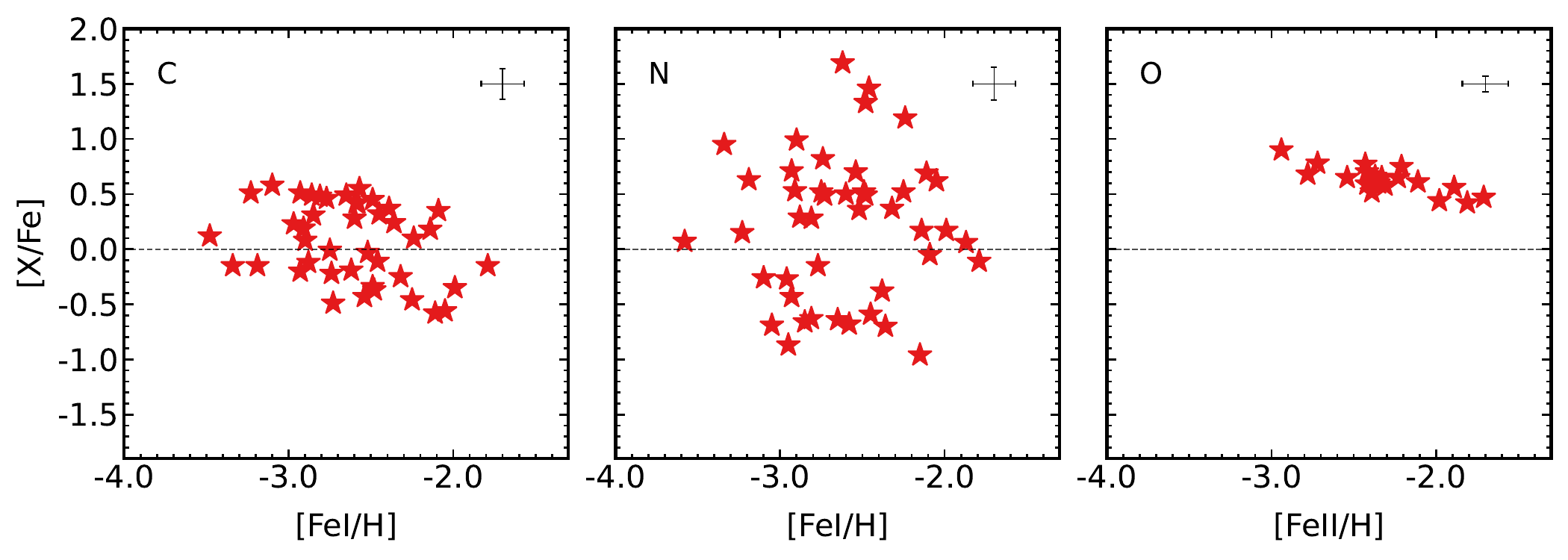}
     \caption{Elemental abundances for C, N, and O plotted as a function of [Fe/H] for the stars in our sample. A representative error bar is plotted in the upper right corner of each panel.}
     \label{fig:cno_feh}
\end{figure*}

\begin{figure}
  \resizebox{\hsize}{!}{\includegraphics{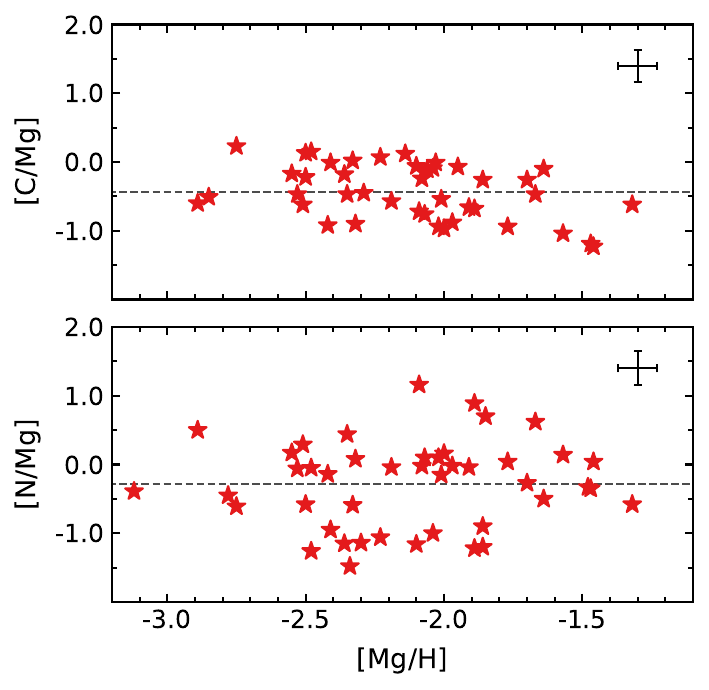}}
  \caption{[C/Mg] and [N/Mg] as a function of [Mg/H] for the stars in our sample. A representative error bar is plotted in the upper-right corner of each panel.}
  \label{fig:cmg_mgh}
\end{figure}

Continuing with the CNO elements, the left panel in Fig.~\ref{fig:cno_feh} shows the derived abundances for carbon compared to \ion{Fe}{i}. The temperature range of the stars in our sample allowed us deriving carbon abundances using the molecular G-band for 42 out of the 52 stars. For the ten remaining stars, carbon measurements are not possible because their spectra do not cover the region between 4277\,\AA\ and 4330\,\AA\ and the atomic \ion{C}{iii} line at 5376.19\,\AA\  is too weak. The mean value of the ratio [C/Fe] is +0.063\RM{~dex} with a scatter of $\sigma =$ 0.366\RM{~dex}. Our carbon abundances span from $\sim-$0.58 to $\sim$+0.58\RM{~dex} and therefore there are no CEMP stars\footnote{[C/Fe]>0.7\RM{~dex} and [Fe/H]$<-2$\RM{~dex} and additionally the Sr/Ba ratio cf. \cite{Hansen2019}} in the sample. This is, however, justified by the fact that we avoided selecting stars with strong carbon molecular features. 
The large scatter observed with respect to Fe is comparable to the one found when plotting [C/Mg] versus [Mg/H] (Fig.~\ref{fig:cmg_mgh}, upper panel).

The middle panel in Fig.~\ref{fig:cno_feh} shows the relation between nitrogen and \ion{Fe}{i}. The mean value $<$[N/Fe]$>$ is 0.21\RM{~dex} and the dispersion is 0.658\RM{~dex}. This value is even greater than that found for C. The same occurs for the ratio [N/Mg] (Fig.~\ref{fig:cmg_mgh}, lower panel) where the dispersion of 0.630\RM{~dex} is almost the same as for [N/Fe]. The large scatter found for [C/Fe] and [N/Fe] versus [Fe/H] may indicate that subsequent mixing episodes have altered the initial composition of the atmosphere in some stars.

\RM{For the star CES\,1245$-$2425, also known as HE\,1243$-$2408, we found an abundance of A(C) = 5.85\,dex. This result is similar to the abundance derived in \cite{2024MNRAS.527.2323S} for the same star, where they found A(C) = 5.95\,dex from spectral synthesis in the CH G-band at 4315\AA\ and before applying evolutionary corrections in order to take into account the extra-mixing. With regards to nitrogen, however, our results show a large discrepancy, which might be due to the different nitrogen lines considered. We obtained an abundance of A(N) = 4.35\,dex using the NH band, whereas \citeauthor{2024MNRAS.527.2323S} obtained an upper limit of A(N)$<$8.33 by using the \element[][12]{C}N line at 4215\AA. 
}

As  noted in Section~\ref{oxygen}, the two neutral oxygen lines used in this work are not affected by NLTE effects, in contrast with the predictions that \RM{ionized} species are less affected by NLTE than the neutral ones \citep{2016MNRAS.463.1518A}. Therefore, we plotted oxygen compared to \ion{Fe}{ii} (right panel of Fig.~\ref{fig:cno_feh}). We find that the ratio [O/Fe] declines linearly \RM{toward} solar metallicities (Fig.~\ref{fig:cno_feh}), with [O/Fe] = 1.01\RM{~dex} at [Fe/H] = $-$3.05\RM{~dex} decreasing to [O/Fe] = 0.63\RM{~dex} at [Fe/H] = $-$1.87\RM{~dex}. The estimated slope is about $-$0.25. The mean value we found for oxygen is [O/Fe] $\sim$ 0.77\RM{~dex} and the dispersion of 0.10\RM{~dex} is the lowest among those calculated for CNO. The average value obtained for oxygen abundances is close to that in \cite{2004A&A...416.1117C} for metal-poor halo stars, where they also used the forbidden oxygen line at 6300\,$\AA$.

\section{Discussion \label{sec:discussion}}

The stars may preserve information on the composition of the gas from which they formed if they have not evolved considerably. Metal-poor stars are of particular interest as these objects are likely a product of the earliest generation of stars formed in the Universe and can thus provide us with archaeological information about the evolution of the early Galaxy. \RM{However, the
dredge-up events can alter their initial composition bringing
material from deep layers to the surface. When the star ascend to RGB, the first dredge-up (FDU) occurs and the abundance of the light elements on the star's surface is predicted to change \citep{1964ApJ...140.1631I, 1984PhR...105..329I}. Although this event is expected to be less efficient at low-metallicities \citep{1994A&A...282..811C}, a second mixing episode takes place when the star becomes brighter than the RGB bump. This extra-mixing is responsible for destroying the remaining Li, as well as depleting $^{12}$C while increasing $^{14}$N, resulting in anomalous abundance patterns for these elements \citep{2000A&A...354..169G}. Therefore, it is important to know if the stars we are analyzing have gone through extra-mixing.}

\begin{figure}
  \resizebox{\hsize}{!}{\includegraphics{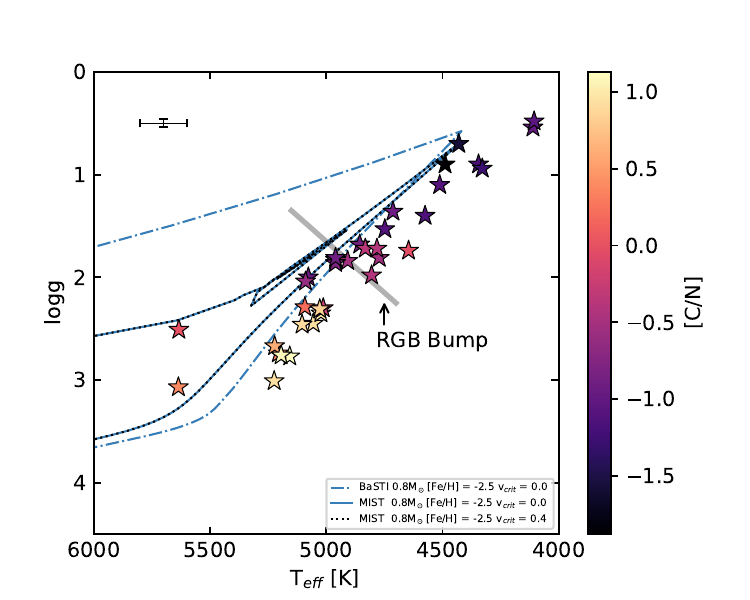}}
  \caption{CERES stars in the \RM{$T_\mathrm{eff}$} - log\,$g$ diagram. The BaSTI evolutionary track (blue dash-dotted line) has an $\alpha$-enhanced chemical mixture and does not include rotation. Two MIST evolutionary tracks are also included for non-rotating (blue continuous line) and with rotation v$_{crit}$ = 0.4 (gray dotted line). All the tracks are for M = 0.8~M$_{\odot}$ and [Fe/H] = $-$2.50\RM{~dex}. The RGB Bump for the \RM{adopted} tracks is at T$_{eff} \sim$ 4880\,K and \RM{$\mathrm{log}\,g$} $\sim$1.95\,dex. The color index on the right side of the figure indicates the [C/N] abundance ratio. The dispersion in \RM{$T_\mathrm{eff}$} and \RM{$\mathrm{log}\,g$} measurements is represented by the error bar in the upper left corner of the figure. 
  }
  \label{fig:hr_diagram_cn}
\end{figure}

The location of our program stars along the \RM{HR diagram} indicates that almost all of them are in the RGB phase (Fig.~\ref{fig:hr_diagram_cn}). The exception is the chemically peculiar star \RM{CES\,2250$-$4057,} which is likely a horizontal branch (HB) star{\footnote{With an absolute $G$ magnitude of 0.28~mag, from its distance modulus of 9.68~mag, and de-reddened $G_{BP}-G_{RP}$ colour of 0.77~mag \citep[Paper~I,][]{2021AJ....161..147B,2021A&A...649A...1G}, using the mean extinction coefficients from \citet[][]{2018MNRAS.479L.102C} for \textit{Gaia} passbands and the reddening map from \citet[][]{2011ApJ...737..103S}, \RM{CES\,2250$-$4057} lies in the region of the CMD occupied by red horizontal branch stars \citep[see, e.g., Fig.~3 from][]{2018A&A...616A..10G}. }}. In Paper\,I, we showed that the ratios [\ion{Y}{ii}/\ion{Sr}{ii}] and [\ion{Zr}{ii}/\ion{Sr}{ii}] for this star \RM{have} the same pattern as other HB stars studied by \cite{2014AJ....147..136R}. \RM{Interestingly, although these results suggest that the star CES\,2250$-$4057 is a HB star,  we found that it shows no sign of extra-mixing.} We have used evolutionary tracks from BaSTI and MIST with [Fe/H] = $-$2.50\RM{~dex}. Both tracks agree with the phase of the sample stars within the scatter and comparison with evolutionary tracks accounting for rotation does not show any detectable difference compared to those with V$_{rot}$= 0 for our low-mass stars. We adopted the models for 0.8-M$_{\odot}$ stars, however, using the tracks with masses ranging from $\sim0.6$ to $\sim1.2$~M$_{\odot}$ does not  significantly change the evolutionary stage of our RGB stars. The \RM{color code} with regard to carbon-nitrogen ratios shows that  the spectroscopically observed [C/N] decreases from the base to the tip of the RGB. It is also important to state that this ratio changes due to both the increasing [N/H] and the decreasing [C/H]. Such abundance changes on the stellar surface have been observed in metal-poor stars and indicate deep mixing taking place in the stars during the RGB phase \citep{2000A&A...354..169G, 2008AJ....136.2522M}.

\subsection{Carbon and nitrogen}

During the CNO cycle, N is produced at the expense of C. The FDU will enrich the surface of the star and turn the atmosphere of the star rich in $^{14}$N and poor in $^{12}$C, relative to its initial composition. As a result, carbon and nitrogen can be used as mixing indicators  \citep[e.g.,][]{2000A&A...354..169G, Korn2006Natur}. Abundance changes in N at the expense of O are not expected because the ON-cycle occurs in a region deeper than the CN-cycle and requires higher temperatures than the temperatures involved in our low-mass stars. 

From our results for carbon and nitrogen shown in Table~\ref{tab:abundances}, it is evident that some of the stars exhibit an enrichment of nitrogen and a depletion of carbon. In fact, the mean average abundance found for nitrogen is relatively higher than that found for carbon. This \RM{behavior} is  seen more clearly when plotting [N/Fe] against [C/Fe] (Fig.~\ref{fig:nfe_cfe}). In spite of this, there is no fully clear separation between the stars that could indicate internal mixing. Although we have considered all the stars in the sample, comprising different metallicities, the plots for given bins of metallicities in the range of $\sim0.5$\,dex (Figs.~\ref{fig:ncfe_wmet} and \ref{fig:cfe_nfe_logg_teff_wmet}) still do not show a clear distinction. By adopting the limits defined by \cite{2005A&A...430..655S}, the distinction between mixed and unmixed stars, however, becomes more clear, especially if we are taking into account the error bars estimated for the stars ($\sigma_{[C/Fe]}$ = $\pm$0.18\RM{~dex} and $\sigma_{[N/Fe]}$ = $\pm$0.22\RM{~dex}). We have found that only two out of the 35 stars for which we obtained both carbon and nitrogen abundances could not be classified according to the limits from \citeauthor{2005A&A...430..655S}. These stars are \RM{CES\,0221$-$2130} with [C/Fe] = $-$0.35\RM{~dex} and [N/Fe] = $+$0.17\RM{~dex} and \RM{CES\,0424$-$1501,} with [C/Fe] = $-$1.94~\RM{dex} and [N/Fe] = $-$0.11~\RM{dex} (blue filled circles in the figures) and are the most metal-rich stars in the sample with carbon and nitrogen measurements. Another four stars are classified as unmixed within the abundance uncertainties (encircled stars). Additionally, we have five more stars with no carbon measurements, which we classified as unmixed due to the presence of Li in their atmospheres. 

When plotting [C/Fe] against the effective temperature and \RM{$\mathrm{log}\,g$} (upper panels in Fig.~\ref{fig:cfe_nfe_teff_logg}), it is noticeable that the sample splits in two distinct groups following our classification. The diagram is color-coded according to the mixing, which shows that there is a plateau around [C/Fe] $\sim$0.4\,dex for the unmixed stars; however, there is an exception for the stars \RM{CES\,0527$-$2052, CES\,1222$-$1136, CES\,1413$-$7609, and CES\,2254$-$4209}, which  are placed along with the mixed stars. These four outliers are the unmixed stars that are the closest to the "mixed region" in Fig.~\ref{fig:nfe_cfe} (circled symbols) and two of them could also be classified as mixed stars considering the estimated uncertainties for both carbon and nitrogen. Moreover, they are located after the RGB bump (yellow dashed lines in Fig.~\ref{fig:cfe_nfe_teff_logg}), \RM{where extra-mixing begins due to thermohaline instability
\citep{2007A&A...467L..15C}.} Hence, these stars are likely to have completed the FDU and have begun to experience extra-mixing, even though carbon depletion and nitrogen enrichment are moderate. Regarding the mixed stars, although they are clearly separated from the flat trend displayed by the unmixed stars, they have a more scattered distribution with no tendency to lower [C/Fe] ratios \RM{toward} lower temperatures -- as  would be expected from deep mixing. %[RGB BUMP LINES - describe here]

In contrast, the plots for [N/Fe] versus \RM{$T_\mathrm{eff}$} and \RM{$\mathrm{log}\,g$} trends (middle panels in Fig.~\ref{fig:cfe_nfe_teff_logg}) do not follow the above-mentioned behavior. Instead, they display a single linear trend for both classes of stars regardless of the outliers. As the trends for N are steeper than the ones observed for carbon, this could indicate that maybe the ON-cycle has taken place and converted some O into N. However, the lower panels show that oxygen remains constant with the temperature and \RM{$\mathrm{log}\,g$} and the fewer unmixed stars observed in the plot do not allow statistically solid inferences.

By plotting C, N, and O in terms of \RM{[X/H]} (see Fig.~\ref{fig:ch_nh_teff_logg}) we obtained a wider spread in the figure, yet the trends with temperature and surface gravity remain. One possible scenario to explain this is that these stars have been formed in clouds with different pre-existing carbon and nitrogen reservoirs and therefore not all nitrogen observed on their surfaces is a product of deep mixing. Another possibility is that this could be evidence of primary N production.

\subsection{Carbon and nitrogen versus oxygen}

Different mechanisms taking place in stars of different mass ranges are responsible for producing the CNO elements. As a consequence, the [C/O] and [N/O] ratios measured in low-metallicity stars in the Galaxy exhibit a remarkable deviation from 0 \citep[see e.g.,][]{2004A&A...416.1117C,2005A&A...430..655S,2014A&A...568A..25N,2019A&A...622L...4A,2019A&A...630A.104A}. The ratios [C/O] and [N/O] can be of particular interest for studying the chemical evolution of galaxies (see Sect.~\ref{sec:GCE}).

\begin{figure}
  \resizebox{\hsize}{!}{\includegraphics{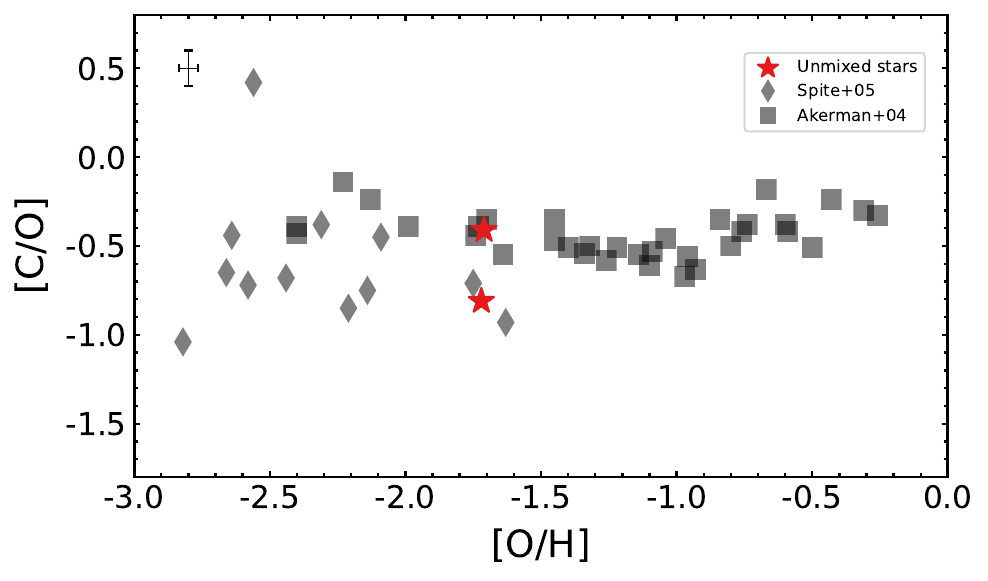}}
  \caption{[C/O] ratios versus [O/H] for the stars in our sample. Gray symbols represent [C/O] ratios derived in \cite{2005A&A...430..655S} (diamond symbols) and \cite{2004A&A...414..931A} (square symbols).}
  \label{fig:coh}
\end{figure}

\begin{figure}
  \resizebox{\hsize}{!}{\includegraphics{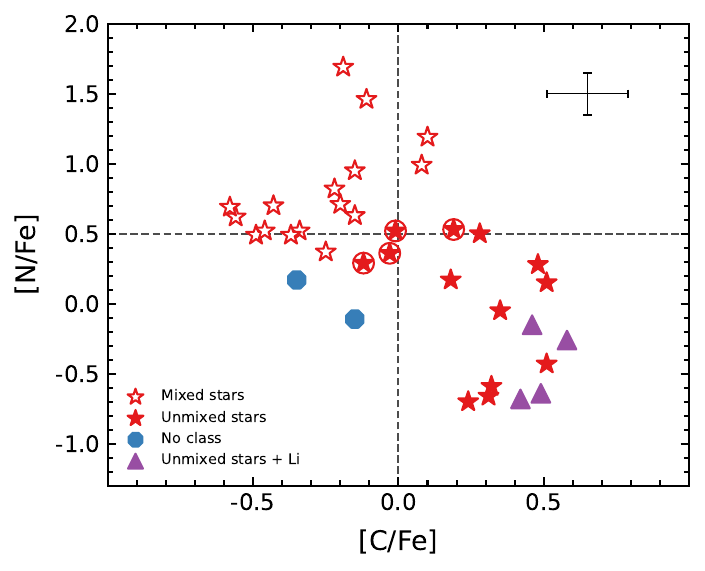}}
  \caption{[N/Fe] versus [C/Fe] for our sample. The mixed and unmixed stars are classified based on \cite{2005A&A...430..655S} and considering the estimated uncertainties for each element. The open red star markers represent the mixed stars, while the filled red star markers represent the unmixed stars; the violet triangles are the unmixed stars for which we also found Li abundances; and the blue hexagons are the stars \RM{CES\,0221$-$2130 and CES\,0424$-$1501}  outside the adopted limits for mixed and unmixed stars. The encircled red stars marker are unmixed stars classified within the abundance uncertainties and are located after the RGB Bump in the \RM{HR diagram}. A representative error bar is plotted in the upper-right corner of the figure.}
  \label{fig:nfe_cfe}
\end{figure}

\begin{figure*}
\centering
   \includegraphics[width=18cm]{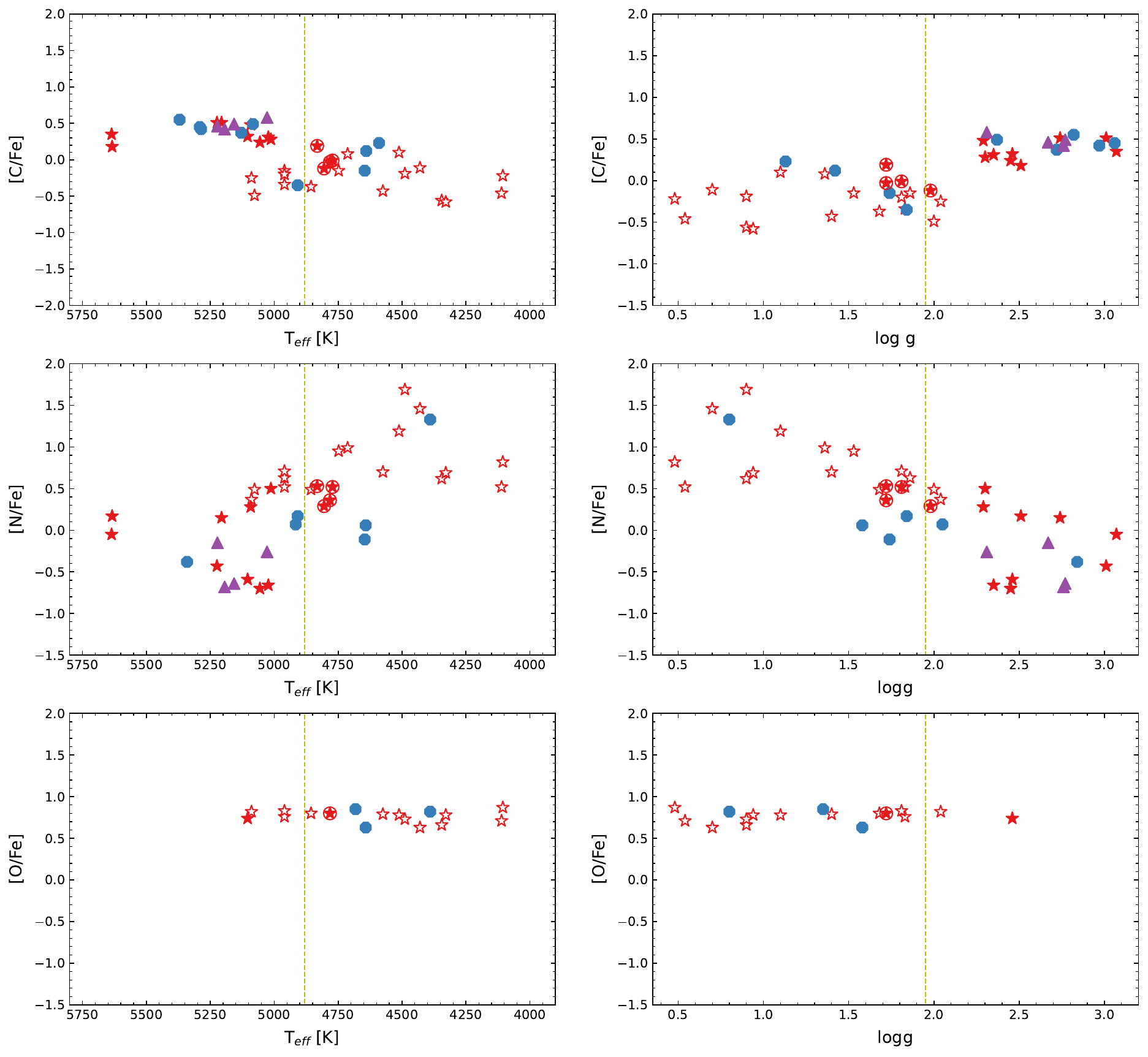}
  \caption{ [C/Fe] as a function of the effective temperature (on the left) and the surface gravity (on the right) for all the stars in our sample (top). A clear separation can be seen between the mixed and unmixed stars, except for the four encircled stars. [N/Fe] with respect to \RM{$T_\mathrm{eff}$} and \RM{$\mathrm{log}\,g$} are shown in the middle panels. The lower panel shows [O/Fe] as a function of \RM{$T_\mathrm{eff}$} and \RM{$\mathrm{log}\,g$}. The yellow dashed line indicates the \RM{$T_\mathrm{eff}$} and \RM{$\mathrm{log}\,g$} at the RGB bump. The blue symbols represent the stars for which we could not classify from C and N, therefore, the blue symbols in the [C/Fe] plots not necessarily are the same stars seen in the plot [N/Fe]. Symbols are the same as in Fig.\ref{fig:nfe_cfe}.}
  \label{fig:cfe_nfe_teff_logg}
\end{figure*}

Figure~\ref{fig:coh} shows [C/O] ratios as a function of [O/H] for both mixed and unmixed stars in our sample. Oxygen is mainly produced in massive stars on short timescale \citep{2006ApJ...653.1145K,2009A&A...505..605C}, so that the plot for [C/O] against [O/H] depends mainly on the yields and timescale of carbon production and can thus offer clues on carbon evolution.
It has been shown that at lower metallicities, the ratio [C/O] increases with decreasing [O/H] \citep{2004A&A...414..931A,2009A&A...500.1143F}. However, the 3D non-LTE analysis \citep{2019A&A...630A.104A} seems to contradict these results. This trend might be connected to the carbon-rich yields from the massive Population III stars \citep{2014A&A...562A.146I}; alternatively, these metal-poor stars could be fast rotators \citep{2006A&A...447..623M, 2006A&A...449L..27C}. Although these observations disagree with the predictions of some chemical evolution models \citep{2003MNRAS.339...63C,2005A&A...432..861G}, slow-rotating stellar models at $Z = 10^{-5}$ have shown that [C/O] ratios might increase with increasing [O/H] \cite{2008A&A...489..685E}. It would be interesting also to investigate [C/O] ratios relatively to [O/H] for the unmixed stars; unfortunately, we have managed to measure oxygen only for two of the stars that fall into the unmixed stars group and any further interpretation is not possible. However, the [C/O] ratios for these two stars are in good agreement with results from literature for halo stars.

The behavior of the ratio [N/O] relative to the metallicity is associated to the nature of nitrogen. Stellar studies at low metallicities have found that the abundance of primary nitrogen remains constant with respect to [O/H] \citep{2004A&A...421..649I, 2005A&A...430..655S} but here we find quite some scatter around the flat trend. However, if the nitrogen production were dominated by secondary mechanisms, one should expect nitrogen abundances proportional to [O/H] ratios \citep{1978MNRAS.185P..77E,2010ApJ...720..226P,2021MNRAS.502.4359R}. In Fig.~\ref{fig:noh}, we compare the [N/O] and [O/H] ratios.
Here, again, any statistical analysis is limited by the small number of [N/O] ratios we obtained for the stars in our sample. Additionally taking into account the stars for which we have Li abundances despite the absence of carbon abundances, we have [N/O] ratios for four unmixed stars. In comparison with results from literature, our stars also show a good agreement, although the two unmixed stars classified from Li abundances have lower [N/O] ratios.  

\subsection{Lithium}
Lithium was primordially synthesized in the Big Bang nucleosynthesis and is easily destroyed at temperatures higher than $~2.5\times10^6$ K. This means that Li can be preserved only in the atmosphere of the stellar surface in evolved stars, not in their interiors. After the main sequence, when the star ascends to the RGB and goes through the FDU, mixing processes are responsible for destroying the Li present on the surface of the star, reducing the observed Li abundance of the star. Therefore, it is also possible to probe deep mixing from Li abundances. Measurements of Li abundances were possible for only nine stars in our sample (Fig.~\ref{fig:li_fe}). All these stars show A(Li) $\geq$ 0.94 and a mean of 1.02 with low dispersion. We were able to prove mixing for four out of these nine stars. According to the adopted limits, these four stars have Li abundances compatible with unmixed stars, namely, [N/Fe] $<$ 0.5\RM{~dex} and [C/N] $>$ $-$0.6\RM{~dex}. For the other five stars, no measurements of carbon or nitrogen were possible. In contrast, we did not detect lithium in any of the mixed stars ([N/Fe] $>$ 0.5\RM{~dex} and [C/N] $<$ $-$0.6\RM{~dex}).

\subsection{Rotation}
\begin{figure}[h!]
  \resizebox{\hsize}{!}{\includegraphics{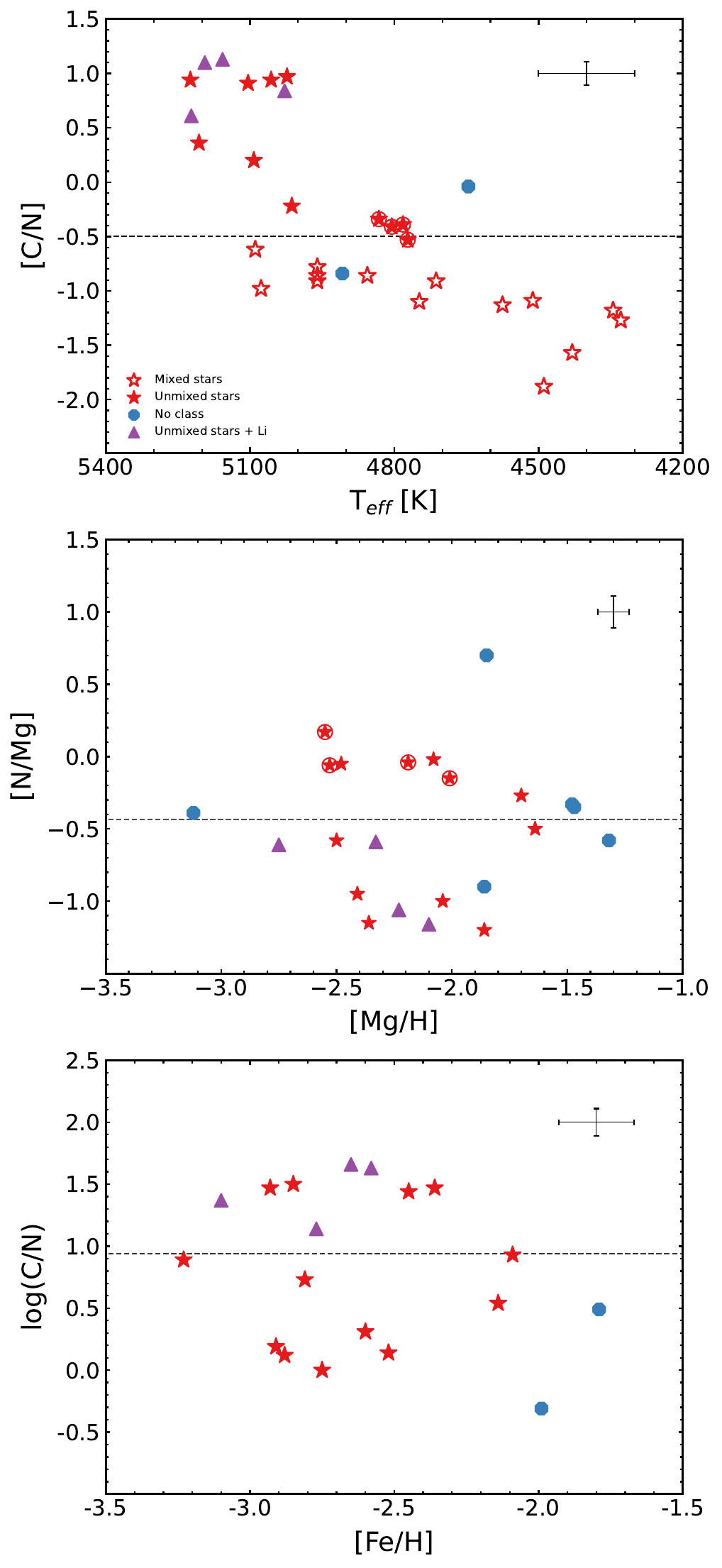}}
  \caption{[C/N] ratios as a function of \RM{$T_\mathrm{eff}$} (top). The dashed line at [C/N] = $-$0.6\RM{~dex} separates the mixed and unmixed stars. Middle panel: [N/Mg] ratios as a function of [Mg/H]. The horizontal dashed line at [N/Mg] = $\sim$0.43\RM{~dex} indicates the mean value for the ratio [N/Mg]. Lower panel: log(C/N) as a function of [Fe/H]. Symbols are the same as in Fig.~\ref{fig:nfe_cfe}. A representative error bar is plotted in the upper-right corner of the panels.}
  \label{fig:nmg_mgh}
\end{figure}

Rotation has been proven to play a key role in reproducing the evolution of massive stars at very \RM{low metallicities} \citep[e.g.,][]{2006A&A...447..623M,2006A&A...449L..27C}. The high rotation rates are responsible for internal mixing, leading to a large production of $^{14}$N and $^{13}$C \citep{2006A&A...447..623M}. As a consequence, higher carbon-to-nitrogen ratios should be expected in non-rotating stars with respect to their counterparts that have gone through rotational mixing. In this sense, if the clouds where the stars in our sample have formed were made up of material enriched by previous rotating models, our unmixed stars (i.e., stars whose surface abundances  directly reflect the abundances of the material from which they were formed) would be expected to show lower C/N ratios than stars made from material enriched by slowly or non-rotating models.
This \RM{behavior} is well reproduced by the stars in our sample as shown in the upper panel of Fig.~\ref{fig:nmg_mgh}. The plot appears to follow a trend to lower [C/N] ratios with decreasing \RM{$T_\mathrm{eff}$}. The stars that fall into the mixed stars group populate the region further down below [C/N] $<$ $-$0.6\RM{~dex,} while the ones in the unmixed stars group have [C/N] ratios above this value. \RM{These results are in good agreement with the predictions of \cite{2017A&A...601A..27L, 2019A&A...621A..24L} using the Besançon Galaxy model, accounting for thermohaline instability effects on the theoretical [C/N] ratios along the RGB. }
Here, we note two things, namely, that we are neglecting possible differences in initial stellar enrichment (e.g., from massive stars with or without rotation) and there is a region between $\sim4800-5100$\,K where the mixed and unmixed groups overlap.

\begin{figure}
  \resizebox{\hsize}{!}{\includegraphics{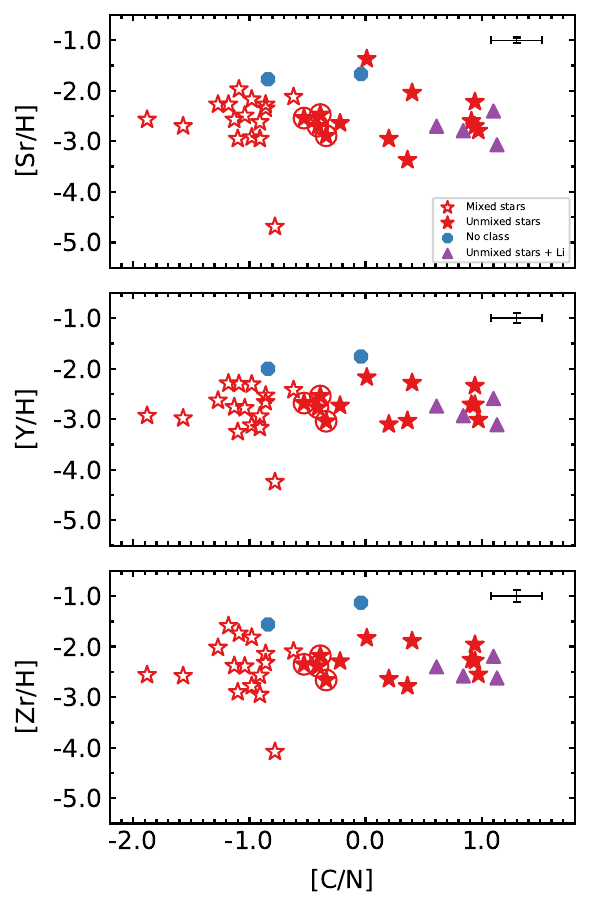}}
  \caption{Light s-process elements Sr, Y, and Zr as a function of [C/N] for the stars in our sample. A representative error bar is plotted in the upper-left corner of the figure. Symbols are indicated in the upper panel of this figure and are the same as in Fig.~\ref{fig:nfe_cfe}.}
  \label{fig:sr_cn}
\end{figure}

\cite{2016ApJ...827..126B,2019ApJ...874...93B} have found that the C/N ratio shows a flat trend at log(C/N) $\sim$ 0.90 (whatever the metallicity) but with a significant scatter ($\sigma \sim$ 0.5). This plateau may indicate that carbon and nitrogen are produced by the same nucleosynthetic mechanisms. In plotting log(C/N) versus [Fe/H] for our unmixed stars (lower panel of Fig.~\ref{fig:nmg_mgh}), we also find that the C/N ratio is almost constant with the metallicity (log(C/N) $\sim$ 0.94) and a scatter of 0.54, close to the observations by Berg et al. 

Beyond the contribution to the light element abundances, the mixing induced by rotation also leads to a great enhancement of $^{22}$Ne \citep{2006A&A...447..623M}. This may, in turn, contribute with neutrons to the weak s-process through the reaction $^{22}$Ne($\alpha$, n)$^{25}$Mg, causing an overproduction of light s-process elements as Sr, Y, and Zr \citep{2011Natur.472..454C, 2012A&A...538L...2F}. However, this does not seem to be the case of our stars (Paper\,I). Moreover, we see no clear trend in Sr, Y, or Zr as a function of [C/H], [N/H], or [C/N] regardless of the mixing or evolutionary stage (see Fig.\ref{fig:sr_cn})\footnote{Linear trends have been fit and their slopes confirm this - see Table~\ref{slopes}}.

\begin{figure}
  \resizebox{\hsize}{!}{\includegraphics{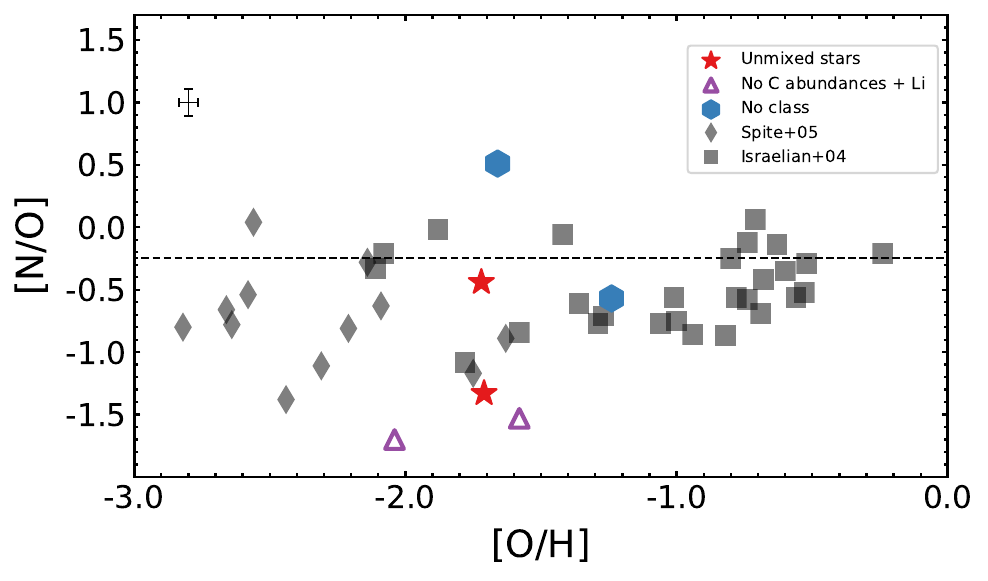}}
  \caption{[N/O] ratios versus [O/H] for the stars in our sample. Gray symbols represent [N/O] ratios derived in \cite{2005A&A...430..655S} (diamond symbols) and \cite{2004A&A...414..931A} (square symbols).}
  \label{fig:noh}
\end{figure}

\subsection{Comparison with chemical evolution model results \label{sec:GCE}}

Hereafter, we present our comparison of  the abundance data given in the previous sections with the predictions of a GCE model. The model assumes that the thick- and thin-disc components of the Milky Way form sequentially, out of two main episodes of gas accretion \citep[two-infall model,][]{1997ApJ...477..765C,2001ApJ...554.1044C,2010A&A...522A..32R,2019A&A...623A..60S}. We adopted the latest version of the model, which includes constraints on stellar ages from asteroseismology \citep[][and references therein]{2019A&A...623A..60S,2021A&A...647A..73S}. In particular, the delay time for the second infall is fixed to a much higher value (3.25~Gyr) than assumed in the original formulation (1~Gyr). The infalling gas is accreted smoothly, following an exponentially decaying law. In the solar vicinity, the e-folding time scales are 1 and 7~Gyr for the thick- and thin-disc components, respectively. The star formation rate is expressed as a \citet{1998ApJ...498..541K} law,
\begin{equation}
\psi(t) \propto \sigma_{\mathrm{gas}}^k(t),
\end{equation}
where $\sigma_{\mathrm{gas}}(t)$ is the surface gas density at any time and $k = 1.5$. The adopted galaxy-wide stellar initial mass function (gwIMF) is that of \citet{1993MNRAS.262..545K}.

The most important ingredients of any GCE model are the stellar yields, though. For the present study, mostly concerned with the chemical composition of metal-poor stars, we fixed the yields of low- and intermediate-mass stars \citep[taken from the FRUITY web-based database,][]{2009ApJ...696..797C,2011ApJS..197...17C,2015ApJS..219...40C} and explore different nucleosynthesis prescriptions for massive stars.

First, we considered the recommended yield set (set R) of \citet{2018ApJS..237...13L}. These authors computed the presupernova evolution and explosive nucleosynthesis of massive star models covering the initial mass range 13--120~M$_\odot$ for different initial chemical compositions (corresponding to [Fe/H]~$= -3$, $-2$, $-1$, and 0) and rotational velocities of the stars ($v_\mathrm{rot}$~= 0, 150, and 300~km~s$^{-1}$). In set R, the mixing and fallback scheme \citep{2002ApJ...565..385U} was adopted for all stars in the initial mass range 13--25~M$_\odot$. The inner border of the mixed zone was set at the layer where [Ni/Fe]~= 0.2\RM{~dex}, while the outer border was fixed at the base of the O shell.
The mass cut is then set by the assumption that each star ejects 0.07 M$_\odot$ of Ni. Stars more massive than 25~M$_\odot$ collapse to black holes; thus, their yields consist of the sole wind contribution. The impact of massive star rotation on the evolution of CNO isotopes predicted by adopting \citet{2018ApJS..237...13L} yields is extensively discussed in \citet{2019MNRAS.490.2838R,2020A&A...639A..37R}. The \RM{light pink}, pink, and violet lines in Fig.~\ref{fig:models} show the predictions of the GCE model implementing \citet{2018ApJS..237...13L} yields for massive stars with $v_\mathrm{rot} =$ 0, 150, and 300~km~s$^{-1}$, respectively. It is seen that the model that does not include fast rotators predicts abundance trends completely offset from the data, due to the well \RM{documented} underproduction of N when rotation is not included in the massive stars models \citep{2006A&A...449L..27C, 2010A&A...522A..32R, 2019MNRAS.490.2838R, 2018MNRAS.476.3432P, 2020ApJ...900..179K}.

The yield set R of \citet{2018ApJS..237...13L} % along with \citet{Limongi2012} and \citet{Roberti2024} 
covers homogeneously an extended range of initial masses, metallicities and rotational velocities, and allows studying a number of nuclear species. 
Other studies in the literature investigate the modifications in stellar nucleosynthesis triggered by rotation, but are limited to the pre-SN stages and/or to a few chemical elements \citep[e.g.,][]{2002A&A...390..561M, 2005A&A...433.1013H, 2007A&A...461..571H, 2008A&A...489..685E}. Nonetheless, these yields have been included in GCE models, providing interesting clues on the origin of $^{14}$N and $^{13}$C in the solar vicinity \citep{2006A&A...449L..27C, 2010A&A...522A..32R} and across the whole Galactic disc \citep{2017MNRAS.470..401R}. 
In this work, we tested the latest generation of stellar models by the Geneva group (using the Geneva stellar evolution code or {\sc Genec}\footnote{A detailed description can be found in \citet{Eggenberger2008}.}), by implementing in the adopted GCE code yields for massive stars, in the mass range of 9~{\msol} $\leqslant M_{\rm ini} \leqslant$ 120~{\msol}, which are either rotating\footnote{They have an initial equatorial velocity of 40\% of the critical one. The critical limit is defined when the gravitational acceleration is counterbalanced by the centrifugal force at the equator \citep{MaederMeynet2000}.} {\RM{(dark green} and dark-blue lines in Fig.~\ref{fig:models}) or non-rotating models (light-green and \RM{light blue} lines in Fig.~\ref{fig:models}). We included stellar grids from \ST{extremely metal-poor} up to super-solar metallicity \citep[more details  in][]{Ekstrom2012, Georgy2012, Groh2019, Sibony2024}.  % Murphy2021a, Eggenberger2021, Yusof2022

To calculate the stellar yields, either during the lifetime or at the end of it, we need to define the mass cut, which is the limit below which everything will be locked in the stellar remnant, while the rest will be ejected to the ISM. We investigate the effects of changing the mass cut location. Thus, we followed two ways of defining the mass cut, either from \citet{Patton2020}, shown as dashed greenish curves in Fig.~\ref{fig:models}, or from \citet{Maeder1992}, shown as dashed blueish curves in Fig.~\ref{fig:models}. For both cases, it is necessary to calculate the CO core mass ($M_{\rm CO}$) of the models, which is defined where the mass fraction of helium drops below 0.01\footnote{A detailed prescription for the first case can be found in \citet{Sibony2024}.}. Having the remnant masses, we obtain the stellar yields by integrating above the mass cut and by subtracting the initial abundance of the isotopes. 
%\RM{[=Concerning, the $^{52}$Fe stellar yields from {\sc Genec} models, since they do not reach the core Si-burning phase, we extrapolated those values using the $^{52}$Fe stellar yields from \citet{2018ApJS..237...13L, Limongi2012} and \citet{Roberti2024} depending the $M_{\rm CO}$.=]} 
It is found, again, that stellar rotation boosts the production of N at low metallicities, which improves the agreement between theoretical predictions and observational data. Moreover, it is found that the adopted mass cut location strongly influences primary N production from rotating massive star models otherwise sharing the same initial rotational velocity.

\ST{In Fig.~\ref{fig:SMC_yields}, we present the stellar yields in $M_{\odot}$ as calculated in Tsiatsiou et al. (in prep.) and used for the GCE models in Fig.\ref{fig:models}. We compare the stellar yields from {\sc Genec} models \citep{Georgy2012} with those from \citet{2018ApJS..237...13L} for the non-rotating models. The two different prescriptions for calculating the remnant masses in the {\sc Genec} models are represented by blue curves \citep{Maeder1992} and green curves \citep{Patton2020}. By comparing non-rotating models, we can eliminate differences due to the physics of rotation.
We note here that in \citet{2018ApJS..237...13L}, for models with initial masses above 25~{\msol}, only the stellar winds are taken into account since these stars are assumed to form black holes that swallow the entire mass of the star at core collapse. The same applies to the {\sc Genec} models from \citet{Maeder1992}. However, in the {\sc Genec} models, when the remnant mass is computed according to \citet{Patton2020}, models with initial masses above 20~{\msol} are assumed to become black holes; however, not all of the final mass of the star is swallowed by the black hole. Significant portions of the star's mass can be lost through the supernova explosion prior to the black hole formation.

By examining Fig.~\ref{fig:models}, we observe that the differences in the stellar yields of the CNO elements between the {\sc Genec} models and the \citet{2018ApJS..237...13L} tracks can be attributed to differences in overshooting parameters, mass loss prescriptions, and the final evolutionary stages. For example, the {\sc Genec} models are not computed until the pre-supernova stage. 
Despite these differences, we see that the yields of $^{12}$C are similar. However, there is a significant exception for the 20~{\msol} model, which shows a very small $^{12}$C yield when the remnant mass is determined using the prescription by \citet{Patton2020}.
Similar values were also obtained for the yields of $^{14}$N for stars with initial masses below about 60~{\msol}. The differences in the mass seen above likely result from variations in the stellar wind prescriptions, especially the metallicity dependence of the stellar winds. % From the mass-loss rates scaling with metallicity: Limongi uses a = 0.85, while Genec a = 0.85 or 0.5 or 0.66
The yields of $^{16}$O are very similar, with the same remark regarding the differences shown by the 20~{\msol} model computed using the prescription by \citet{Patton2020} for the determination of the remnant mass.}

%The models by \citet{2018ApJS..237...13L} use an overshooting parameter of $\alpha_{\rm ov}=0.2$, while the {\sc Genec} models use $\alpha_{\rm ov}=0.1$. The {\sc Genec} models tend to produce higher yields for C and O at certain masses, while the \citet{2018ApJS..237...13L} tracks show higher N yields at very high initial masses. These variations underscore the sensitivity of stellar nucleosynthesis to model parameters and evolutionary processes.}

\begin{figure}
  \resizebox{\hsize}{!}{\includegraphics{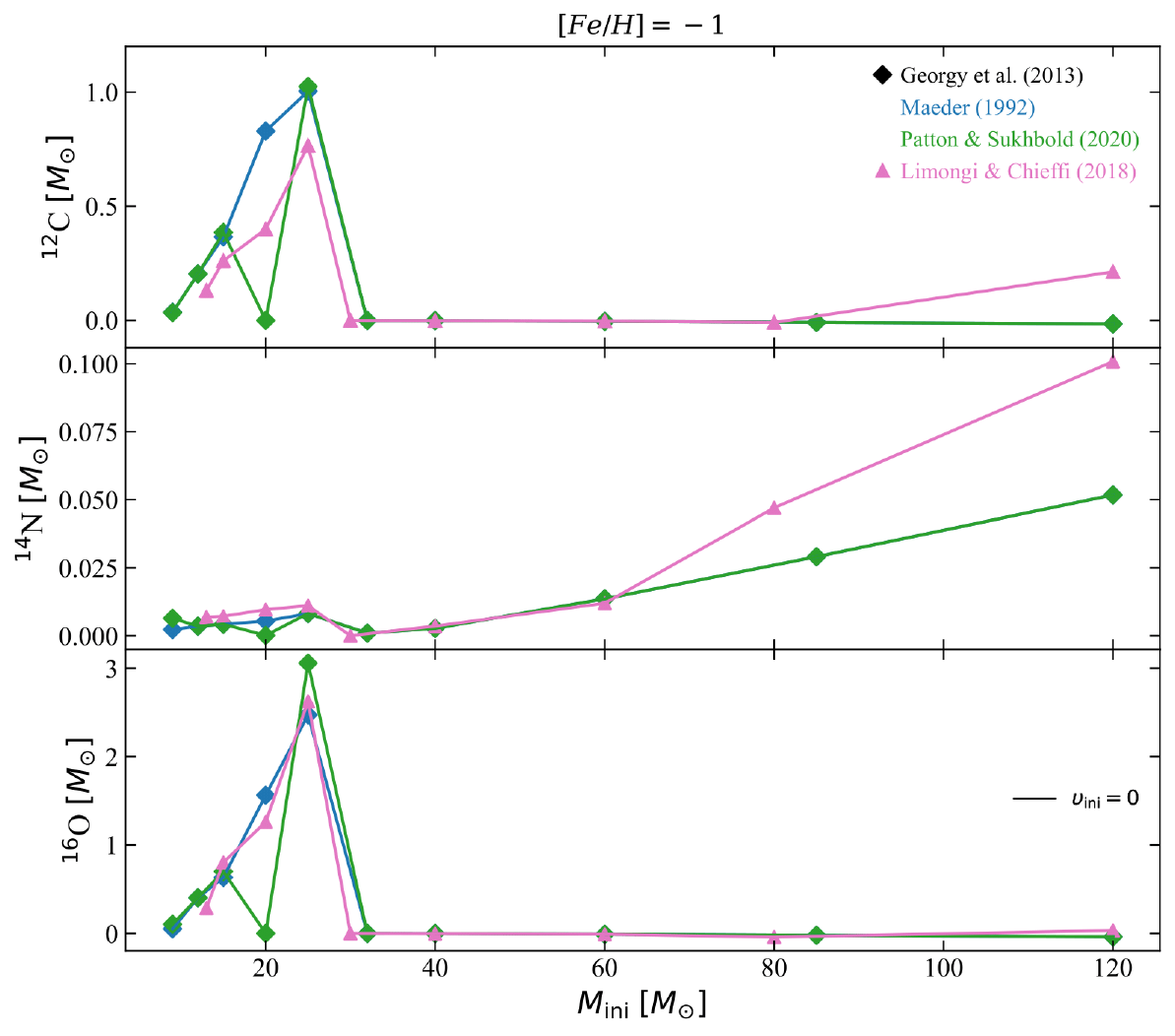}}
  \caption{\ST{Stellar yields in $M_{\odot}$ for CNO elements for non-rotating models with $[{\rm Fe}/{\rm H}]=-1$, from {\sc Genec} models \citep[diamonds, ][]{Georgy2012} and \citet{2018ApJS..237...13L} (triangles, pink curves). The models have an initial mass range of 9~{\msol} $\leqslant M_{\rm ini} \leqslant$ 120~{\msol}. The blue curves of {\sc Genec} models correspond to the \citet{Maeder1992} prescription of calculating the remnant mass, and the green curves to the \citet{Patton2020}.}}
  \label{fig:SMC_yields}
\end{figure}

\section{Conclusions \label{sec:conclusion}}

In this work, we have performed a detailed chemical analysis of the surface abundances of carbon, nitrogen, oxygen, and lithium in a sample of 52 halo stars with metallicities in the range $-$3.49 $\leq$[Fe/H] $\leq$ $-$1.79\RM{~dex}. Our individual abundances presented here were computed under the 1D LTE assumption. The 3D NLTE corrections for lithium calculated by \citet{2021MNRAS.500.2159W} did not significantly impact  our results.

The star \RM{CES\,2250$-$4057} is the only star in our sample lying in the region associated with the horizontal branch. It has been shown in Paper\,I that this is a striking star with an overabundance of Sr with respect to Y and Zr, while all abundances are low on a relative scale. In addition, we found that \RM{this star} has mild carbon and nitrogen enrichment with [C/Fe] = 0.18\RM{~dex} and [N/Fe] = 0.17\RM{~dex} and, thus, does not show traces of deep mixing according to our classification.

Our RGB stars can be split in three different groups.
The group of mixed stars have an overabundance of nitrogen and carbon depletion showing clear signs that the CN-cycle processed material was brought to the surface of these stars. Additionally, we found no signature that could indicate that the ON-cycle took place. According to the temperature and surface gravities of these stars, these stars are on the upper RGB, after the RGB Bump, showing deep mixing and confirming the results on extra-mixing \citep{2000A&A...354..169G}.

The stars classified as unmixed \citep[see][]{2005A&A...430..655S} show no signature of CNO processed material at their surfaces and exhibit [C/N] ratios lower than $-$0.50\RM{~dex}. The presence of lithium on the surface of nine of the unmixed stars was also used to probe the mixing (and the abundance values confirm that they are unmixed). Almost all these stars lie on the lower RGB with temperatures and surface \RM{$\mathrm{log}\,g$} above $\sim$4880\,K and $\sim$1.95\,dex, respectively. However, for four unmixed stars, we found that they approach the upper RGB and their [N/Fe] ratios are close to our adopted limits. This may indicate that the signature of mixing due to the CNO burning layers is not yet significant. 

The last group consists of stars that we did not manage to classify given the lack of carbon or nitrogen and also lithium. These stars show no particular behavior when plotted along with the other stars of the sample and (according to their position on the \RM{HR diagram}) it is likely that some of them have experienced deep mixing.
When considering heavy elements Sr, Y, and Zr, neither of the three mixing groups show specific trends as a function of [C/N]. This may indicate that these heavy elements do not suffer detectable changes in the stellar surface for stars with our sample stellar parameters. It is important to know that the nucleosynthetic imprints of heavy elements remain unaltered by stellar evolution processes and the abundances can directly be used to trace the underlying neutron capture processes. 

We computed homogeneous GCE models to trace the evolution of carbon, nitrogen, and oxygen in our Galaxy. We confirm that models taking stellar rotation into account fit the average runs of abundance ratios for unmixed stars better than models that do not consider the effects of rotation. We also find that the location of the mass cut influences the exact amount of primary N ejected by low-metallicity stars, at a fixed initial rotational velocity.

\begin{figure*}[h!]
  \resizebox{\hsize}{!}{\includegraphics{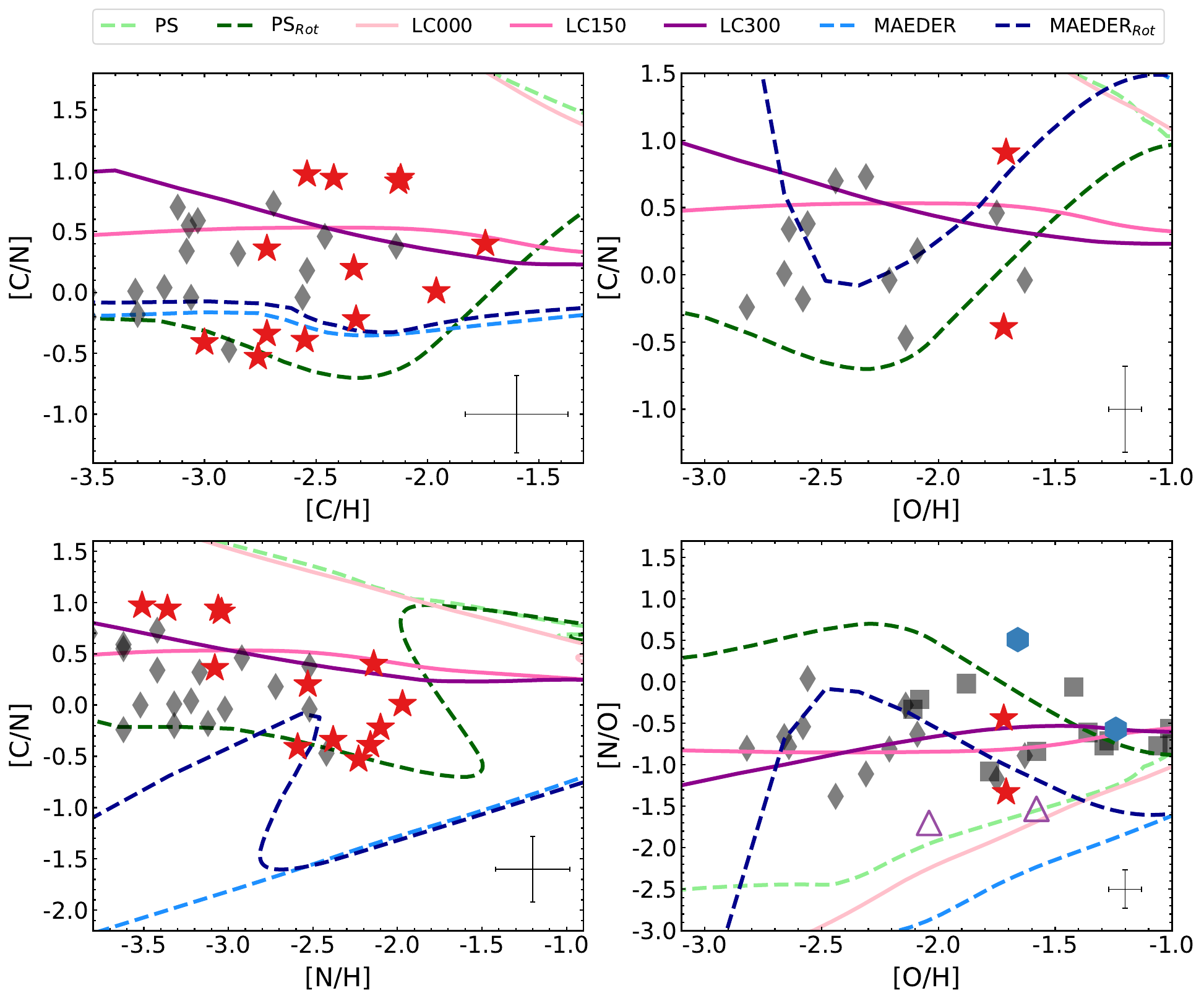}}
  \caption{[C/N] ratios as a function of [C/H] (top left panel), [O/H] (top right panel), and [N/H] (bottom left panel) for the unmixed stars in our sample. The \RM{bottom} right panel shows [N/O] ratios as a function of [O/H] for the unmixed stars in our sample. Abundance ratios from \citeauthor{2005A&A...430..655S} (gray diamond symbols) and \citeauthor{2004A&A...421..649I} (gray square symbols) for low-metallicity stars are also shown in the figures. \RM{The continuous lines are the predictions of GCE models using \cite{2018ApJS..237...13L} yields for massive stars with rotational velocities of v$_{rot}$ = 0 (LC000), 150\,km\,s$^{-1}$ (LC150), and 300\,km\,s$^{-1}$ (LC300); whereas the dashed lines correspond to GCE model predictions obtained using yields from GENEC stellar models with \cite{Patton2020} and \cite{Maeder1992} mass cut prescriptions for non-rotating (PS and MAEDER, respectively) and rotating stars (PS$_{rot}$ and MAEDER$_{Rot}$, respectively).} The models are color-coded as indicated in the top of the figure. A representative error bar is plotted in the lower right corner of each panel.}
  \label{fig:models}
\end{figure*}

\begin{acknowledgements}
\RM{We thank the anonymous referee for the useful comments and suggestions that helped to improve this work.} RFM, AAP, LL, and CJH acknowledge the support by the State of Hesse within the Research Cluster ELEMENTS (Project ID 500/10.006). CJH also acknowledges the European Union’s Horizon 2020 research and innovation \RM{program} under grant agreement No 101008324 (ChETEC-INFRA). ST and GM have received funding from the European Research Council (ERC) under the European Union's Horizon 2020 research and innovation \RM{program} (grant agreement No 833925, project STAREX). DR acknowledges the support by the Italian National Institute for Astrophysics (INAF) through Finanziamento della Ricerca Fondamentale, Theory Grant Fu.~Ob.~1.05.12.06.08 (Project ``An in-depth theoretical study of CNO element evolution in galaxies'').
\end{acknowledgements}


%\begin{thebibliography}{}

%  \bibitem[Baker(1966)]{baker} Baker, N. 1966,
%      in Stellar Evolution,
%      ed.\ R. F. Stein,\& A. G. W. Cameron
%      (Plenum, New York) 333

%\end{thebibliography}
\bibliographystyle{aa_url} % style aa.bst
\bibliography{cno} % your references Yourfile.bib

\begin{thebibliography}{104}
\expandafter\ifx\csname natexlab\endcsname\relax\def\natexlab#1{#1}\fi

\bibitem[{{Akerman} {et~al.}(2004){Akerman}, {Carigi}, {Nissen}, {Pettini}, \& {Asplund}}]{2004A&A...414..931A}
{Akerman}, C.~J., {Carigi}, L., {Nissen}, P.~E., {Pettini}, M., \& {Asplund}, M. 2004, \href{http://dx.doi.org/10.1051/0004-6361:20034188}{\color{magenta}\aap}, \href{https://ui.adsabs.harvard.edu/abs/2004A&A...414..931A}{414, 931}

\bibitem[{{Amarsi} {et~al.}(2016{\natexlab{a}}){Amarsi}, {Asplund}, {Collet}, \& {Leenaarts}}]{Amarsi2016MNRAS.455.3735A}
{Amarsi}, A.~M., {Asplund}, M., {Collet}, R., \& {Leenaarts}, J. 2016{\natexlab{a}}, \href{http://dx.doi.org/10.1093/mnras/stv2608}{\color{magenta}\mnras}, \href{https://ui.adsabs.harvard.edu/abs/2016MNRAS.455.3735A}{455, 3735}

\bibitem[{{Amarsi} {et~al.}(2016{\natexlab{b}}){Amarsi}, {Lind}, {Asplund}, {Barklem}, \& {Collet}}]{2016MNRAS.463.1518A}
{Amarsi}, A.~M., {Lind}, K., {Asplund}, M., {Barklem}, P.~S., \& {Collet}, R. 2016{\natexlab{b}}, \href{http://dx.doi.org/10.1093/mnras/stw2077}{\color{magenta}\mnras}, \href{https://ui.adsabs.harvard.edu/abs/2016MNRAS.463.1518A}{463, 1518}

\bibitem[{{Amarsi} {et~al.}(2019{\natexlab{a}}){Amarsi}, {Nissen}, {Asplund}, {Lind}, \& {Barklem}}]{2019A&A...622L...4A}
{Amarsi}, A.~M., {Nissen}, P.~E., {Asplund}, M., {Lind}, K., \& {Barklem}, P.~S. 2019{\natexlab{a}}, \href{http://dx.doi.org/10.1051/0004-6361/201834480}{\color{magenta}\aap}, \href{https://ui.adsabs.harvard.edu/abs/2019A&A...622L...4A}{622, L4}

\bibitem[{{Amarsi} {et~al.}(2019{\natexlab{b}}){Amarsi}, {Nissen}, \& {Sk{\'u}lad{\'o}ttir}}]{2019A&A...630A.104A}
{Amarsi}, A.~M., {Nissen}, P.~E., \& {Sk{\'u}lad{\'o}ttir}, {\'A}. 2019{\natexlab{b}}, \href{http://dx.doi.org/10.1051/0004-6361/201936265}{\color{magenta}\aap}, \href{https://ui.adsabs.harvard.edu/abs/2019A&A...630A.104A}{630, A104}

\bibitem[{{Arcones} \& {Thielemann}(2023)}]{2023A&ARv..31....1A}
{Arcones}, A. \& {Thielemann}, F.-K. 2023, \href{http://dx.doi.org/10.1007/s00159-022-00146-x}{\color{magenta}\aapr}, \href{https://ui.adsabs.harvard.edu/abs/2023A&ARv..31....1A}{31, 1}

\bibitem[{{Bailer-Jones} {et~al.}(2021){Bailer-Jones}, {Rybizki}, {Fouesneau}, {Demleitner}, \& {Andrae}}]{2021AJ....161..147B}
{Bailer-Jones}, C.~A.~L., {Rybizki}, J., {Fouesneau}, M., {Demleitner}, M., \& {Andrae}, R. 2021, \href{http://dx.doi.org/10.3847/1538-3881/abd806}{\color{magenta}\aj}, \href{https://ui.adsabs.harvard.edu/abs/2021AJ....161..147B}{161, 147}

\bibitem[{{Berg} {et~al.}(2019){Berg}, {Erb}, {Henry}, {Skillman}, \& {McQuinn}}]{2019ApJ...874...93B}
{Berg}, D.~A., {Erb}, D.~K., {Henry}, R. B.~C., {Skillman}, E.~D., \& {McQuinn}, K. B.~W. 2019, \href{http://dx.doi.org/10.3847/1538-4357/ab020a}{\color{magenta}\apj}, \href{https://ui.adsabs.harvard.edu/abs/2019ApJ...874...93B}{874, 93}

\bibitem[{{Berg} {et~al.}(2016){Berg}, {Skillman}, {Henry}, {Erb}, \& {Carigi}}]{2016ApJ...827..126B}
{Berg}, D.~A., {Skillman}, E.~D., {Henry}, R. B.~C., {Erb}, D.~K., \& {Carigi}, L. 2016, \href{http://dx.doi.org/10.3847/0004-637X/827/2/126}{\color{magenta}\apj}, \href{https://ui.adsabs.harvard.edu/abs/2016ApJ...827..126B}{827, 126}

\bibitem[{{Burbidge} {et~al.}(1957){Burbidge}, {Burbidge}, {Fowler}, \& {Hoyle}}]{1957RvMP...29..547B}
{Burbidge}, E.~M., {Burbidge}, G.~R., {Fowler}, W.~A., \& {Hoyle}, F. 1957, \href{http://dx.doi.org/10.1103/RevModPhys.29.547}{\color{magenta}Reviews of Modern Physics}, \href{https://ui.adsabs.harvard.edu/abs/1957RvMP...29..547B}{29, 547}

\bibitem[{{Caffau} {et~al.}(2011){Caffau}, {Faraggiana}, {Ludwig}, {Bonifacio}, \& {Steffen}}]{2011AN....332..128C}
{Caffau}, E., {Faraggiana}, R., {Ludwig}, H.~G., {Bonifacio}, P., \& {Steffen}, M. 2011, \href{http://dx.doi.org/10.1002/asna.201011485}{\color{magenta}Astronomische Nachrichten}, \href{https://ui.adsabs.harvard.edu/abs/2011AN....332..128C}{332, 128}

\bibitem[{{Casagrande} \& {VandenBerg}(2018)}]{2018MNRAS.479L.102C}
{Casagrande}, L. \& {VandenBerg}, D.~A. 2018, \href{http://dx.doi.org/10.1093/mnrasl/sly104}{\color{magenta}\mnras}, \href{https://ui.adsabs.harvard.edu/abs/2018MNRAS.479L.102C}{479, L102}

\bibitem[{{Cayrel} {et~al.}(2004){Cayrel}, {Depagne}, {Spite}, {Hill}, {Spite}, {Fran{\c{c}}ois}, {Plez}, {Beers}, {Primas}, {Andersen}, {Barbuy}, {Bonifacio}, {Molaro}, \& {Nordstr{\"o}m}}]{2004A&A...416.1117C}
{Cayrel}, R., {Depagne}, E., {Spite}, M., {et~al.} 2004, \href{http://dx.doi.org/10.1051/0004-6361:20034074}{\color{magenta}\aap}, \href{https://ui.adsabs.harvard.edu/abs/2004A&A...416.1117C}{416, 1117}

\bibitem[{{Cescutti} {et~al.}(2009){Cescutti}, {Matteucci}, {McWilliam}, \& {Chiappini}}]{2009A&A...505..605C}
{Cescutti}, G., {Matteucci}, F., {McWilliam}, A., \& {Chiappini}, C. 2009, \href{http://dx.doi.org/10.1051/0004-6361/200912759}{\color{magenta}\aap}, \href{https://ui.adsabs.harvard.edu/abs/2009A&A...505..605C}{505, 605}

\bibitem[{{Charbonnel}(1994)}]{1994A&A...282..811C}
{Charbonnel}, C. 1994, \aap, \href{https://ui.adsabs.harvard.edu/abs/1994A&A...282..811C}{282, 811}

\bibitem[{{Charbonnel} \& {Zahn}(2007)}]{2007A&A...467L..15C}
{Charbonnel}, C. \& {Zahn}, J.~P. 2007, \href{http://dx.doi.org/10.1051/0004-6361:20077274}{\color{magenta}\aap}, \href{https://ui.adsabs.harvard.edu/abs/2007A&A...467L..15C}{467, L15}

\bibitem[{{Chiappini} {et~al.}(2011){Chiappini}, {Frischknecht}, {Meynet}, {Hirschi}, {Barbuy}, {Pignatari}, {Decressin}, \& {Maeder}}]{2011Natur.472..454C}
{Chiappini}, C., {Frischknecht}, U., {Meynet}, G., {et~al.} 2011, \href{http://dx.doi.org/10.1038/nature10000}{\color{magenta}\nat}, \href{https://ui.adsabs.harvard.edu/abs/2011Natur.472..454C}{472, 454}

\bibitem[{{Chiappini} {et~al.}(2006){Chiappini}, {Hirschi}, {Meynet}, {Ekstr{\"o}m}, {Maeder}, \& {Matteucci}}]{2006A&A...449L..27C}
{Chiappini}, C., {Hirschi}, R., {Meynet}, G., {et~al.} 2006, \href{http://dx.doi.org/10.1051/0004-6361:20064866}{\color{magenta}\aap}, \href{https://ui.adsabs.harvard.edu/abs/2006A&A...449L..27C}{449, L27}

\bibitem[{{Chiappini} {et~al.}(1997){Chiappini}, {Matteucci}, \& {Gratton}}]{1997ApJ...477..765C}
{Chiappini}, C., {Matteucci}, F., \& {Gratton}, R. 1997, \href{http://dx.doi.org/10.1086/303726}{\color{magenta}\apj}, \href{https://ui.adsabs.harvard.edu/abs/1997ApJ...477..765C}{477, 765}

\bibitem[{{Chiappini} {et~al.}(2001){Chiappini}, {Matteucci}, \& {Romano}}]{2001ApJ...554.1044C}
{Chiappini}, C., {Matteucci}, F., \& {Romano}, D. 2001, \href{http://dx.doi.org/10.1086/321427}{\color{magenta}\apj}, \href{https://ui.adsabs.harvard.edu/abs/2001ApJ...554.1044C}{554, 1044}

\bibitem[{{Chiappini} {et~al.}(2003){Chiappini}, {Romano}, \& {Matteucci}}]{2003MNRAS.339...63C}
{Chiappini}, C., {Romano}, D., \& {Matteucci}, F. 2003, \href{http://dx.doi.org/10.1046/j.1365-8711.2003.06154.x}{\color{magenta}\mnras}, \href{https://ui.adsabs.harvard.edu/abs/2003MNRAS.339...63C}{339, 63}

\bibitem[{{Collet} {et~al.}(2007){Collet}, {Asplund}, \& {Trampedach}}]{2007A&A...469..687C}
{Collet}, R., {Asplund}, M., \& {Trampedach}, R. 2007, \href{http://dx.doi.org/10.1051/0004-6361:20066321}{\color{magenta}\aap}, \href{https://ui.adsabs.harvard.edu/abs/2007A&A...469..687C}{469, 687}

\bibitem[{{Cristallo} {et~al.}(2011){Cristallo}, {Piersanti}, {Straniero}, {Gallino}, {Dom{\'\i}nguez}, {Abia}, {Di Rico}, {Quintini}, \& {Bisterzo}}]{2011ApJS..197...17C}
{Cristallo}, S., {Piersanti}, L., {Straniero}, O., {et~al.} 2011, \href{http://dx.doi.org/10.1088/0067-0049/197/2/17}{\color{magenta}\apjs}, \href{https://ui.adsabs.harvard.edu/abs/2011ApJS..197...17C}{197, 17}

\bibitem[{{Cristallo} {et~al.}(2009){Cristallo}, {Straniero}, {Gallino}, {Piersanti}, {Dom{\'\i}nguez}, \& {Lederer}}]{2009ApJ...696..797C}
{Cristallo}, S., {Straniero}, O., {Gallino}, R., {et~al.} 2009, \href{http://dx.doi.org/10.1088/0004-637X/696/1/797}{\color{magenta}\apj}, \href{https://ui.adsabs.harvard.edu/abs/2009ApJ...696..797C}{696, 797}

\bibitem[{{Cristallo} {et~al.}(2015){Cristallo}, {Straniero}, {Piersanti}, \& {Gobrecht}}]{2015ApJS..219...40C}
{Cristallo}, S., {Straniero}, O., {Piersanti}, L., \& {Gobrecht}, D. 2015, \href{http://dx.doi.org/10.1088/0067-0049/219/2/40}{\color{magenta}\apjs}, \href{https://ui.adsabs.harvard.edu/abs/2015ApJS..219...40C}{219, 40}

\bibitem[{{Dekker} {et~al.}(2000){Dekker}, {D'Odorico}, {Kaufer}, {Delabre}, \& {Kotzlowski}}]{2000SPIE.4008..534D}
{Dekker}, H., {D'Odorico}, S., {Kaufer}, A., {Delabre}, B., \& {Kotzlowski}, H. 2000, in Society of Photo-Optical Instrumentation Engineers (SPIE) Conference Series, Vol. 4008, Optical and IR Telescope Instrumentation and Detectors, ed. M.~{Iye} \& A.~F. {Moorwood}, \href{https://ui.adsabs.harvard.edu/abs/2000SPIE.4008..534D}{534--545}

\bibitem[{{Edmunds} \& {Pagel}(1978)}]{1978MNRAS.185P..77E}
{Edmunds}, M.~G. \& {Pagel}, B.~E.~J. 1978, \href{http://dx.doi.org/10.1093/mnras/185.1.77P}{\color{magenta}\mnras}, \href{https://ui.adsabs.harvard.edu/abs/1978MNRAS.185P..77E}{185, 77P}

\bibitem[{{Eggenberger} {et~al.}(2008){Eggenberger}, {Meynet}, {Maeder}, {Hirschi}, {Charbonnel}, {Talon}, \& {Ekstr{\"o}m}}]{Eggenberger2008}
{Eggenberger}, P., {Meynet}, G., {Maeder}, A., {et~al.} 2008, \href{http://dx.doi.org/10.1007/s10509-007-9511-y}{\color{magenta}\apss}, \href{https://ui.adsabs.harvard.edu/abs/2008Ap&SS.316...43E}{316, 43}

\bibitem[{{Ekstr{\"o}m} {et~al.}(2012){Ekstr{\"o}m}, {Georgy}, {Eggenberger}, {Meynet}, {Mowlavi}, {Wyttenbach}, {Granada}, {Decressin}, {Hirschi}, {Frischknecht}, {Charbonnel}, \& {Maeder}}]{Ekstrom2012}
{Ekstr{\"o}m}, S., {Georgy}, C., {Eggenberger}, P., {et~al.} 2012, \href{http://dx.doi.org/10.1051/0004-6361/201117751}{\color{magenta}\aap}, \href{https://ui.adsabs.harvard.edu/abs/2012A&A...537A.146E}{537, A146}

\bibitem[{{Ekstr{\"o}m} {et~al.}(2008){Ekstr{\"o}m}, {Meynet}, {Chiappini}, {Hirschi}, \& {Maeder}}]{2008A&A...489..685E}
{Ekstr{\"o}m}, S., {Meynet}, G., {Chiappini}, C., {Hirschi}, R., \& {Maeder}, A. 2008, \href{http://dx.doi.org/10.1051/0004-6361:200809633}{\color{magenta}\aap}, \href{https://ui.adsabs.harvard.edu/abs/2008A&A...489..685E}{489, 685}

\bibitem[{{Fabbian} {et~al.}(2009){Fabbian}, {Nissen}, {Asplund}, {Pettini}, \& {Akerman}}]{2009A&A...500.1143F}
{Fabbian}, D., {Nissen}, P.~E., {Asplund}, M., {Pettini}, M., \& {Akerman}, C. 2009, \href{http://dx.doi.org/10.1051/0004-6361/200810095}{\color{magenta}\aap}, \href{https://ui.adsabs.harvard.edu/abs/2009A&A...500.1143F}{500, 1143}

\bibitem[{{Farmer} {et~al.}(2021){Farmer}, {Laplace}, {de Mink}, \& {Justham}}]{2021ApJ...923..214F}
{Farmer}, R., {Laplace}, E., {de Mink}, S.~E., \& {Justham}, S. 2021, \href{http://dx.doi.org/10.3847/1538-4357/ac2f44}{\color{magenta}\apj}, \href{https://ui.adsabs.harvard.edu/abs/2021ApJ...923..214F}{923, 214}

\bibitem[{{Frischknecht} {et~al.}(2012){Frischknecht}, {Hirschi}, \& {Thielemann}}]{2012A&A...538L...2F}
{Frischknecht}, U., {Hirschi}, R., \& {Thielemann}, F.~K. 2012, \href{http://dx.doi.org/10.1051/0004-6361/201117794}{\color{magenta}\aap}, \href{https://ui.adsabs.harvard.edu/abs/2012A&A...538L...2F}{538, L2}

\bibitem[{{Gaia Collaboration} {et~al.}(2018){Gaia Collaboration}, {Babusiaux}, {van Leeuwen}, {Barstow}, {Jordi}, {Vallenari}, {Bossini}, {Bressan}, {Cantat-Gaudin}, {van Leeuwen}, {Brown}, {Prusti}, {de Bruijne}, {Bailer-Jones}, {Biermann}, {Evans}, {Eyer}, {Jansen}, {Klioner}, {Lammers}, {Lindegren}, {Luri}, {Mignard}, {Panem}, {Pourbaix}, {Randich}, {Sartoretti}, {Siddiqui}, {Soubiran}, {Walton}, {Arenou}, {Bastian}, {Cropper}, {Drimmel}, {Katz}, {Lattanzi}, {Bakker}, {Cacciari}, {Casta{\~n}eda}, {Chaoul}, {Cheek}, {De Angeli}, {Fabricius}, {Guerra}, {Holl}, {Masana}, {Messineo}, {Mowlavi}, {Nienartowicz}, {Panuzzo}, {Portell}, {Riello}, {Seabroke}, {Tanga}, {Th{\'e}venin}, {Gracia-Abril}, {Comoretto}, {Garcia-Reinaldos}, {Teyssier}, {Altmann}, {Andrae}, {Audard}, {Bellas-Velidis}, {Benson}, {Berthier}, {Blomme}, {Burgess}, {Busso}, {Carry}, {Cellino}, {Clementini}, {Clotet}, {Creevey}, {Davidson}, {De Ridder}, {Delchambre}, {Dell'Oro}, {Ducourant}, {Fern{\'a}ndez-Hern{\'a}ndez}, {Fouesneau},
  {Fr{\'e}mat}, {Galluccio}, {Garc{\'\i}a-Torres}, {Gonz{\'a}lez-N{\'u}{\~n}ez}, {Gonz{\'a}lez-Vidal}, {Gosset}, {Guy}, {Halbwachs}, {Hambly}, {Harrison}, {Hern{\'a}ndez}, {Hestroffer}, {Hodgkin}, {Hutton}, {Jasniewicz}, {Jean-Antoine-Piccolo}, {Jordan}, {Korn}, {Krone-Martins}, {Lanzafame}, {Lebzelter}, {L{\"o}ffler}, {Manteiga}, {Marrese}, {Mart{\'\i}n-Fleitas}, {Moitinho}, {Mora}, {Muinonen}, {Osinde}, {Pancino}, {Pauwels}, {Petit}, {Recio-Blanco}, {Richards}, {Rimoldini}, {Robin}, {Sarro}, {Siopis}, {Smith}, {Sozzetti}, {S{\"u}veges}, {Torra}, {van Reeven}, {Abbas}, {Abreu Aramburu}, {Accart}, {Aerts}, {Altavilla}, {{\'A}lvarez}, {Alvarez}, {Alves}, {Anderson}, {Andrei}, {Anglada Varela}, {Antiche}, {Antoja}, {Arcay}, {Astraatmadja}, {Bach}, {Baker}, {Balaguer-N{\'u}{\~n}ez}, {Balm}, {Barache}, {Barata}, {Barbato}, {Barblan}, {Barklem}, {Barrado}, {Barros}, {Bartholom{\'e} Mu{\~n}oz}, {Bassilana}, {Becciani}, {Bellazzini}, {Berihuete}, {Bertone}, {Bianchi}, {Bienaym{\'e}}, {Blanco-Cuaresma}, {Boch},
  {Boeche}, {Bombrun}, {Borrachero}, {Bouquillon}, {Bourda}, {Bragaglia}, {Bramante}, {Breddels}, {Brouillet}, {Br{\"u}semeister}, {Brugaletta}, {Bucciarelli}, {Burlacu}, {Busonero}, {Butkevich}, {Buzzi}, {Caffau}, {Cancelliere}, {Cannizzaro}, {Carballo}, {Carlucci}, {Carrasco}, {Casamiquela}, {Castellani}, {Castro-Ginard}, {Charlot}, {Chemin}, {Chiavassa}, {Cocozza}, {Costigan}, {Cowell}, {Crifo}, {Crosta}, {Crowley}, {Cuypers}, {Dafonte}, {Damerdji}, {Dapergolas}, {David}, {David}, {de Laverny}, {De Luise}, {De March}, {de Martino}, {de Souza}, {de Torres}, {Debosscher}, {del Pozo}, {Delbo}, {Delgado}, {Delgado}, {Diakite}, {Diener}, {Distefano}, {Dolding}, {Drazinos}, {Dur{\'a}n}, {Edvardsson}, {Enke}, {Eriksson}, {Esquej}, {Eynard Bontemps}, {Fabre}, {Fabrizio}, {Faigler}, {Falc{\~a}o}, {Farr{\`a}s Casas}, {Federici}, {Fedorets}, {Fernique}, {Figueras}, {Filippi}, {Findeisen}, {Fonti}, {Fraile}, {Fraser}, {Fr{\'e}zouls}, {Gai}, {Galleti}, {Garabato}, {Garc{\'\i}a-Sedano}, {Garofalo}, {Garralda}, {Gavel},
  {Gavras}, {Gerssen}, {Geyer}, {Giacobbe}, {Gilmore}, {Girona}, {Giuffrida}, {Glass}, {Gomes}, {Granvik}, {Gueguen}, {Guerrier}, {Guiraud}, {Guti{\'e}}, {Haigron}, {Hatzidimitriou}, {Hauser}, {Haywood}, {Heiter}, {Helmi}, {Heu}, {Hilger}, {Hobbs}, {Hofmann}, {Holland}, {Huckle}, {Hypki}, {Icardi}, {Jan{\ss}en}, {Jevardat de Fombelle}, {Jonker}, {Juh{\'a}sz}, {Julbe}, {Karampelas}, {Kewley}, {Klar}, {Kochoska}, {Kohley}, {Kolenberg}, {Kontizas}, {Kontizas}, {Koposov}, {Kordopatis}, {Kostrzewa-Rutkowska}, {Koubsky}, {Lambert}, {Lanza}, {Lasne}, {Lavigne}, {Le Fustec}, {Le Poncin-Lafitte}, {Lebreton}, {Leccia}, {Leclerc}, {Lecoeur-Taibi}, {Lenhardt}, {Leroux}, {Liao}, {Licata}, {Lindstr{\o}m}, {Lister}, {Livanou}, {Lobel}, {L{\'o}pez}, {Managau}, {Mann}, {Mantelet}, {Marchal}, {Marchant}, {Marconi}, {Marinoni}, {Marschalk{\'o}}, {Marshall}, {Martino}, {Marton}, {Mary}, {Massari}, {Matijevi{\v{c}}}, {Mazeh}, {McMillan}, {Messina}, {Michalik}, {Millar}, {Molina}, {Molinaro}, {Moln{\'a}r}, {Montegriffo}, {Mor},
  {Morbidelli}, {Morel}, {Morris}, {Mulone}, {Muraveva}, {Musella}, {Nelemans}, {Nicastro}, {Noval}, {O'Mullane}, {Ord{\'e}novic}, {Ord{\'o}{\~n}ez-Blanco}, {Osborne}, {Pagani}, {Pagano}, {Pailler}, {Palacin}, {Palaversa}, {Panahi}, {Pawlak}, {Piersimoni}, {Pineau}, {Plachy}, {Plum}, {Poggio}, {Poujoulet}, {Pr{\v{s}}a}, {Pulone}, {Racero}, {Ragaini}, {Rambaux}, {Ramos-Lerate}, {Regibo}, {Reyl{\'e}}, {Riclet}, {Ripepi}, {Riva}, {Rivard}, {Rixon}, {Roegiers}, {Roelens}, {Romero-G{\'o}mez}, {Rowell}, {Royer}, {Ruiz-Dern}, {Sadowski}, {Sagrist{\`a} Sell{\'e}s}, {Sahlmann}, {Salgado}, {Salguero}, {Sanna}, {Santana-Ros}, {Sarasso}, {Savietto}, {Schultheis}, {Sciacca}, {Segol}, {Segovia}, {S{\'e}gransan}, {Shih}, {Siltala}, {Silva}, {Smart}, {Smith}, {Solano}, {Solitro}, {Sordo}, {Soria Nieto}, {Souchay}, {Spagna}, {Spoto}, {Stampa}, {Steele}, {Steidelm{\"u}ller}, {Stephenson}, {Stoev}, {Suess}, {Surdej}, {Szabados}, {Szegedi-Elek}, {Tapiador}, {Taris}, {Tauran}, {Taylor}, {Teixeira}, {Terrett}, {Teyssandier},
  {Thuillot}, {Titarenko}, {Torra Clotet}, {Turon}, {Ulla}, {Utrilla}, {Uzzi}, {Vaillant}, {Valentini}, {Valette}, {van Elteren}, {Van Hemelryck}, {Vaschetto}, {Vecchiato}, {Veljanoski}, {Viala}, {Vicente}, {Vogt}, {von Essen}, {Voss}, {Votruba}, {Voutsinas}, {Walmsley}, {Weiler}, {Wertz}, {Wevers}, {Wyrzykowski}, {Yoldas}, {{\v{Z}}erjal}, {Ziaeepour}, {Zorec}, {Zschocke}, {Zucker}, {Zurbach}, \& {Zwitter}}]{2018A&A...616A..10G}
{Gaia Collaboration}, {Babusiaux}, C., {van Leeuwen}, F., {et~al.} 2018, \href{http://dx.doi.org/10.1051/0004-6361/201832843}{\color{magenta}\aap}, \href{https://ui.adsabs.harvard.edu/abs/2018A&A...616A..10G}{616, A10}

\bibitem[{{Gaia Collaboration} {et~al.}(2021){Gaia Collaboration}, {Brown}, {Vallenari}, {Prusti}, {de Bruijne}, {Babusiaux}, {Biermann}, {Creevey}, {Evans}, {Eyer}, {Hutton}, {Jansen}, {Jordi}, {Klioner}, {Lammers}, {Lindegren}, {Luri}, {Mignard}, {Panem}, {Pourbaix}, {Randich}, {Sartoretti}, {Soubiran}, {Walton}, {Arenou}, {Bailer-Jones}, {Bastian}, {Cropper}, {Drimmel}, {Katz}, {Lattanzi}, {van Leeuwen}, {Bakker}, {Cacciari}, {Casta{\~n}eda}, {De Angeli}, {Ducourant}, {Fabricius}, {Fouesneau}, {Fr{\'e}mat}, {Guerra}, {Guerrier}, {Guiraud}, {Jean-Antoine Piccolo}, {Masana}, {Messineo}, {Mowlavi}, {Nicolas}, {Nienartowicz}, {Pailler}, {Panuzzo}, {Riclet}, {Roux}, {Seabroke}, {Sordo}, {Tanga}, {Th{\'e}venin}, {Gracia-Abril}, {Portell}, {Teyssier}, {Altmann}, {Andrae}, {Bellas-Velidis}, {Benson}, {Berthier}, {Blomme}, {Brugaletta}, {Burgess}, {Busso}, {Carry}, {Cellino}, {Cheek}, {Clementini}, {Damerdji}, {Davidson}, {Delchambre}, {Dell'Oro}, {Fern{\'a}ndez-Hern{\'a}ndez}, {Galluccio}, {Garc{\'\i}a-Lario},
  {Garcia-Reinaldos}, {Gonz{\'a}lez-N{\'u}{\~n}ez}, {Gosset}, {Haigron}, {Halbwachs}, {Hambly}, {Harrison}, {Hatzidimitriou}, {Heiter}, {Hern{\'a}ndez}, {Hestroffer}, {Hodgkin}, {Holl}, {Jan{\ss}en}, {Jevardat de Fombelle}, {Jordan}, {Krone-Martins}, {Lanzafame}, {L{\"o}ffler}, {Lorca}, {Manteiga}, {Marchal}, {Marrese}, {Moitinho}, {Mora}, {Muinonen}, {Osborne}, {Pancino}, {Pauwels}, {Petit}, {Recio-Blanco}, {Richards}, {Riello}, {Rimoldini}, {Robin}, {Roegiers}, {Rybizki}, {Sarro}, {Siopis}, {Smith}, {Sozzetti}, {Ulla}, {Utrilla}, {van Leeuwen}, {van Reeven}, {Abbas}, {Abreu Aramburu}, {Accart}, {Aerts}, {Aguado}, {Ajaj}, {Altavilla}, {{\'A}lvarez}, {{\'A}lvarez Cid-Fuentes}, {Alves}, {Anderson}, {Anglada Varela}, {Antoja}, {Audard}, {Baines}, {Baker}, {Balaguer-N{\'u}{\~n}ez}, {Balbinot}, {Balog}, {Barache}, {Barbato}, {Barros}, {Barstow}, {Bartolom{\'e}}, {Bassilana}, {Bauchet}, {Baudesson-Stella}, {Becciani}, {Bellazzini}, {Bernet}, {Bertone}, {Bianchi}, {Blanco-Cuaresma}, {Boch}, {Bombrun}, {Bossini},
  {Bouquillon}, {Bragaglia}, {Bramante}, {Breedt}, {Bressan}, {Brouillet}, {Bucciarelli}, {Burlacu}, {Busonero}, {Butkevich}, {Buzzi}, {Caffau}, {Cancelliere}, {C{\'a}novas}, {Cantat-Gaudin}, {Carballo}, {Carlucci}, {Carnerero}, {Carrasco}, {Casamiquela}, {Castellani}, {Castro-Ginard}, {Castro Sampol}, {Chaoul}, {Charlot}, {Chemin}, {Chiavassa}, {Cioni}, {Comoretto}, {Cooper}, {Cornez}, {Cowell}, {Crifo}, {Crosta}, {Crowley}, {Dafonte}, {Dapergolas}, {David}, {David}, {de Laverny}, {De Luise}, {De March}, {De Ridder}, {de Souza}, {de Teodoro}, {de Torres}, {del Peloso}, {del Pozo}, {Delbo}, {Delgado}, {Delgado}, {Delisle}, {Di Matteo}, {Diakite}, {Diener}, {Distefano}, {Dolding}, {Eappachen}, {Edvardsson}, {Enke}, {Esquej}, {Fabre}, {Fabrizio}, {Faigler}, {Fedorets}, {Fernique}, {Fienga}, {Figueras}, {Fouron}, {Fragkoudi}, {Fraile}, {Franke}, {Gai}, {Garabato}, {Garcia-Gutierrez}, {Garc{\'\i}a-Torres}, {Garofalo}, {Gavras}, {Gerlach}, {Geyer}, {Giacobbe}, {Gilmore}, {Girona}, {Giuffrida}, {Gomel}, {Gomez},
  {Gonzalez-Santamaria}, {Gonz{\'a}lez-Vidal}, {Granvik}, {Guti{\'e}rrez-S{\'a}nchez}, {Guy}, {Hauser}, {Haywood}, {Helmi}, {Hidalgo}, {Hilger}, {H{\l}adczuk}, {Hobbs}, {Holland}, {Huckle}, {Jasniewicz}, {Jonker}, {Juaristi Campillo}, {Julbe}, {Karbevska}, {Kervella}, {Khanna}, {Kochoska}, {Kontizas}, {Kordopatis}, {Korn}, {Kostrzewa-Rutkowska}, {Kruszy{\'n}ska}, {Lambert}, {Lanza}, {Lasne}, {Le Campion}, {Le Fustec}, {Lebreton}, {Lebzelter}, {Leccia}, {Leclerc}, {Lecoeur-Taibi}, {Liao}, {Licata}, {Lindstr{\o}m}, {Lister}, {Livanou}, {Lobel}, {Madrero Pardo}, {Managau}, {Mann}, {Marchant}, {Marconi}, {Marcos Santos}, {Marinoni}, {Marocco}, {Marshall}, {Martin Polo}, {Mart{\'\i}n-Fleitas}, {Masip}, {Massari}, {Mastrobuono-Battisti}, {Mazeh}, {McMillan}, {Messina}, {Michalik}, {Millar}, {Mints}, {Molina}, {Molinaro}, {Moln{\'a}r}, {Montegriffo}, {Mor}, {Morbidelli}, {Morel}, {Morris}, {Mulone}, {Munoz}, {Muraveva}, {Murphy}, {Musella}, {Noval}, {Ord{\'e}novic}, {Orr{\`u}}, {Osinde}, {Pagani}, {Pagano},
  {Palaversa}, {Palicio}, {Panahi}, {Pawlak}, {Pe{\~n}alosa Esteller}, {Penttil{\"a}}, {Piersimoni}, {Pineau}, {Plachy}, {Plum}, {Poggio}, {Poretti}, {Poujoulet}, {Pr{\v{s}}a}, {Pulone}, {Racero}, {Ragaini}, {Rainer}, {Raiteri}, {Rambaux}, {Ramos}, {Ramos-Lerate}, {Re Fiorentin}, {Regibo}, {Reyl{\'e}}, {Ripepi}, {Riva}, {Rixon}, {Robichon}, {Robin}, {Roelens}, {Rohrbasser}, {Romero-G{\'o}mez}, {Rowell}, {Royer}, {Rybicki}, {Sadowski}, {Sagrist{\`a} Sell{\'e}s}, {Sahlmann}, {Salgado}, {Salguero}, {Samaras}, {Sanchez Gimenez}, {Sanna}, {Santove{\~n}a}, {Sarasso}, {Schultheis}, {Sciacca}, {Segol}, {Segovia}, {S{\'e}gransan}, {Semeux}, {Shahaf}, {Siddiqui}, {Siebert}, {Siltala}, {Slezak}, {Smart}, {Solano}, {Solitro}, {Souami}, {Souchay}, {Spagna}, {Spoto}, {Steele}, {Steidelm{\"u}ller}, {Stephenson}, {S{\"u}veges}, {Szabados}, {Szegedi-Elek}, {Taris}, {Tauran}, {Taylor}, {Teixeira}, {Thuillot}, {Tonello}, {Torra}, {Torra}, {Turon}, {Unger}, {Vaillant}, {van Dillen}, {Vanel}, {Vecchiato}, {Viala}, {Vicente},
  {Voutsinas}, {Weiler}, {Wevers}, {Wyrzykowski}, {Yoldas}, {Yvard}, {Zhao}, {Zorec}, {Zucker}, {Zurbach}, \& {Zwitter}}]{2021A&A...649A...1G}
{Gaia Collaboration}, {Brown}, A.~G.~A., {Vallenari}, A., {et~al.} 2021, \href{http://dx.doi.org/10.1051/0004-6361/202039657}{\color{magenta}\aap}, \href{https://ui.adsabs.harvard.edu/abs/2021A&A...649A...1G}{649, A1}

\bibitem[{{Gaia Collaboration} {et~al.}(2016){Gaia Collaboration}, {Prusti}, {de Bruijne}, {Brown}, {Vallenari}, {Babusiaux}, {Bailer-Jones}, {Bastian}, {Biermann}, {Evans}, {Eyer}, {Jansen}, {Jordi}, {Klioner}, {Lammers}, {Lindegren}, {Luri}, {Mignard}, {Milligan}, {Panem}, {Poinsignon}, {Pourbaix}, {Randich}, {Sarri}, {Sartoretti}, {Siddiqui}, {Soubiran}, {Valette}, {van Leeuwen}, {Walton}, {Aerts}, {Arenou}, {Cropper}, {Drimmel}, {H{\o}g}, {Katz}, {Lattanzi}, {O'Mullane}, {Grebel}, {Holland}, {Huc}, {Passot}, {Bramante}, {Cacciari}, {Casta{\~n}eda}, {Chaoul}, {Cheek}, {De Angeli}, {Fabricius}, {Guerra}, {Hern{\'a}ndez}, {Jean-Antoine-Piccolo}, {Masana}, {Messineo}, {Mowlavi}, {Nienartowicz}, {Ord{\'o}{\~n}ez-Blanco}, {Panuzzo}, {Portell}, {Richards}, {Riello}, {Seabroke}, {Tanga}, {Th{\'e}venin}, {Torra}, {Els}, {Gracia-Abril}, {Comoretto}, {Garcia-Reinaldos}, {Lock}, {Mercier}, {Altmann}, {Andrae}, {Astraatmadja}, {Bellas-Velidis}, {Benson}, {Berthier}, {Blomme}, {Busso}, {Carry}, {Cellino}, {Clementini},
  {Cowell}, {Creevey}, {Cuypers}, {Davidson}, {De Ridder}, {de Torres}, {Delchambre}, {Dell'Oro}, {Ducourant}, {Fr{\'e}mat}, {Garc{\'\i}a-Torres}, {Gosset}, {Halbwachs}, {Hambly}, {Harrison}, {Hauser}, {Hestroffer}, {Hodgkin}, {Huckle}, {Hutton}, {Jasniewicz}, {Jordan}, {Kontizas}, {Korn}, {Lanzafame}, {Manteiga}, {Moitinho}, {Muinonen}, {Osinde}, {Pancino}, {Pauwels}, {Petit}, {Recio-Blanco}, {Robin}, {Sarro}, {Siopis}, {Smith}, {Smith}, {Sozzetti}, {Thuillot}, {van Reeven}, {Viala}, {Abbas}, {Abreu Aramburu}, {Accart}, {Aguado}, {Allan}, {Allasia}, {Altavilla}, {{\'A}lvarez}, {Alves}, {Anderson}, {Andrei}, {Anglada Varela}, {Antiche}, {Antoja}, {Ant{\'o}n}, {Arcay}, {Atzei}, {Ayache}, {Bach}, {Baker}, {Balaguer-N{\'u}{\~n}ez}, {Barache}, {Barata}, {Barbier}, {Barblan}, {Baroni}, {Barrado y Navascu{\'e}s}, {Barros}, {Barstow}, {Becciani}, {Bellazzini}, {Bellei}, {Bello Garc{\'\i}a}, {Belokurov}, {Bendjoya}, {Berihuete}, {Bianchi}, {Bienaym{\'e}}, {Billebaud}, {Blagorodnova}, {Blanco-Cuaresma}, {Boch},
  {Bombrun}, {Borrachero}, {Bouquillon}, {Bourda}, {Bouy}, {Bragaglia}, {Breddels}, {Brouillet}, {Br{\"u}semeister}, {Bucciarelli}, {Budnik}, {Burgess}, {Burgon}, {Burlacu}, {Busonero}, {Buzzi}, {Caffau}, {Cambras}, {Campbell}, {Cancelliere}, {Cantat-Gaudin}, {Carlucci}, {Carrasco}, {Castellani}, {Charlot}, {Charnas}, {Charvet}, {Chassat}, {Chiavassa}, {Clotet}, {Cocozza}, {Collins}, {Collins}, {Costigan}, {Crifo}, {Cross}, {Crosta}, {Crowley}, {Dafonte}, {Damerdji}, {Dapergolas}, {David}, {David}, {De Cat}, {de Felice}, {de Laverny}, {De Luise}, {De March}, {de Martino}, {de Souza}, {Debosscher}, {del Pozo}, {Delbo}, {Delgado}, {Delgado}, {di Marco}, {Di Matteo}, {Diakite}, {Distefano}, {Dolding}, {Dos Anjos}, {Drazinos}, {Dur{\'a}n}, {Dzigan}, {Ecale}, {Edvardsson}, {Enke}, {Erdmann}, {Escolar}, {Espina}, {Evans}, {Eynard Bontemps}, {Fabre}, {Fabrizio}, {Faigler}, {Falc{\~a}o}, {Farr{\`a}s Casas}, {Faye}, {Federici}, {Fedorets}, {Fern{\'a}ndez-Hern{\'a}ndez}, {Fernique}, {Fienga}, {Figueras}, {Filippi},
  {Findeisen}, {Fonti}, {Fouesneau}, {Fraile}, {Fraser}, {Fuchs}, {Furnell}, {Gai}, {Galleti}, {Galluccio}, {Garabato}, {Garc{\'\i}a-Sedano}, {Gar{\'e}}, {Garofalo}, {Garralda}, {Gavras}, {Gerssen}, {Geyer}, {Gilmore}, {Girona}, {Giuffrida}, {Gomes}, {Gonz{\'a}lez-Marcos}, {Gonz{\'a}lez-N{\'u}{\~n}ez}, {Gonz{\'a}lez-Vidal}, {Granvik}, {Guerrier}, {Guillout}, {Guiraud}, {G{\'u}rpide}, {Guti{\'e}rrez-S{\'a}nchez}, {Guy}, {Haigron}, {Hatzidimitriou}, {Haywood}, {Heiter}, {Helmi}, {Hobbs}, {Hofmann}, {Holl}, {Holland}, {Hunt}, {Hypki}, {Icardi}, {Irwin}, {Jevardat de Fombelle}, {Jofr{\'e}}, {Jonker}, {Jorissen}, {Julbe}, {Karampelas}, {Kochoska}, {Kohley}, {Kolenberg}, {Kontizas}, {Koposov}, {Kordopatis}, {Koubsky}, {Kowalczyk}, {Krone-Martins}, {Kudryashova}, {Kull}, {Bachchan}, {Lacoste-Seris}, {Lanza}, {Lavigne}, {Le Poncin-Lafitte}, {Lebreton}, {Lebzelter}, {Leccia}, {Leclerc}, {Lecoeur-Taibi}, {Lemaitre}, {Lenhardt}, {Leroux}, {Liao}, {Licata}, {Lindstr{\o}m}, {Lister}, {Livanou}, {Lobel}, {L{\"o}ffler},
  {L{\'o}pez}, {Lopez-Lozano}, {Lorenz}, {Loureiro}, {MacDonald}, {Magalh{\~a}es Fernandes}, {Managau}, {Mann}, {Mantelet}, {Marchal}, {Marchant}, {Marconi}, {Marie}, {Marinoni}, {Marrese}, {Marschalk{\'o}}, {Marshall}, {Mart{\'\i}n-Fleitas}, {Martino}, {Mary}, {Matijevi{\v{c}}}, {Mazeh}, {McMillan}, {Messina}, {Mestre}, {Michalik}, {Millar}, {Miranda}, {Molina}, {Molinaro}, {Molinaro}, {Moln{\'a}r}, {Moniez}, {Montegriffo}, {Monteiro}, {Mor}, {Mora}, {Morbidelli}, {Morel}, {Morgenthaler}, {Morley}, {Morris}, {Mulone}, {Muraveva}, {Musella}, {Narbonne}, {Nelemans}, {Nicastro}, {Noval}, {Ord{\'e}novic}, {Ordieres-Mer{\'e}}, {Osborne}, {Pagani}, {Pagano}, {Pailler}, {Palacin}, {Palaversa}, {Parsons}, {Paulsen}, {Pecoraro}, {Pedrosa}, {Pentik{\"a}inen}, {Pereira}, {Pichon}, {Piersimoni}, {Pineau}, {Plachy}, {Plum}, {Poujoulet}, {Pr{\v{s}}a}, {Pulone}, {Ragaini}, {Rago}, {Rambaux}, {Ramos-Lerate}, {Ranalli}, {Rauw}, {Read}, {Regibo}, {Renk}, {Reyl{\'e}}, {Ribeiro}, {Rimoldini}, {Ripepi}, {Riva}, {Rixon},
  {Roelens}, {Romero-G{\'o}mez}, {Rowell}, {Royer}, {Rudolph}, {Ruiz-Dern}, {Sadowski}, {Sagrist{\`a} Sell{\'e}s}, {Sahlmann}, {Salgado}, {Salguero}, {Sarasso}, {Savietto}, {Schnorhk}, {Schultheis}, {Sciacca}, {Segol}, {Segovia}, {Segransan}, {Serpell}, {Shih}, {Smareglia}, {Smart}, {Smith}, {Solano}, {Solitro}, {Sordo}, {Soria Nieto}, {Souchay}, {Spagna}, {Spoto}, {Stampa}, {Steele}, {Steidelm{\"u}ller}, {Stephenson}, {Stoev}, {Suess}, {S{\"u}veges}, {Surdej}, {Szabados}, {Szegedi-Elek}, {Tapiador}, {Taris}, {Tauran}, {Taylor}, {Teixeira}, {Terrett}, {Tingley}, {Trager}, {Turon}, {Ulla}, {Utrilla}, {Valentini}, {van Elteren}, {Van Hemelryck}, {van Leeuwen}, {Varadi}, {Vecchiato}, {Veljanoski}, {Via}, {Vicente}, {Vogt}, {Voss}, {Votruba}, {Voutsinas}, {Walmsley}, {Weiler}, {Weingrill}, {Werner}, {Wevers}, {Whitehead}, {Wyrzykowski}, {Yoldas}, {{\v{Z}}erjal}, {Zucker}, {Zurbach}, {Zwitter}, {Alecu}, {Allen}, {Allende Prieto}, {Amorim}, {Anglada-Escud{\'e}}, {Arsenijevic}, {Azaz}, {Balm}, {Beck}, {Bernstein},
  {Bigot}, {Bijaoui}, {Blasco}, {Bonfigli}, {Bono}, {Boudreault}, {Bressan}, {Brown}, {Brunet}, {Bunclark}, {Buonanno}, {Butkevich}, {Carret}, {Carrion}, {Chemin}, {Ch{\'e}reau}, {Corcione}, {Darmigny}, {de Boer}, {de Teodoro}, {de Zeeuw}, {Delle Luche}, {Domingues}, {Dubath}, {Fodor}, {Fr{\'e}zouls}, {Fries}, {Fustes}, {Fyfe}, {Gallardo}, {Gallegos}, {Gardiol}, {Gebran}, {Gomboc}, {G{\'o}mez}, {Grux}, {Gueguen}, {Heyrovsky}, {Hoar}, {Iannicola}, {Isasi Parache}, {Janotto}, {Joliet}, {Jonckheere}, {Keil}, {Kim}, {Klagyivik}, {Klar}, {Knude}, {Kochukhov}, {Kolka}, {Kos}, {Kutka}, {Lainey}, {LeBouquin}, {Liu}, {Loreggia}, {Makarov}, {Marseille}, {Martayan}, {Martinez-Rubi}, {Massart}, {Meynadier}, {Mignot}, {Munari}, {Nguyen}, {Nordlander}, {Ocvirk}, {O'Flaherty}, {Olias Sanz}, {Ortiz}, {Osorio}, {Oszkiewicz}, {Ouzounis}, {Palmer}, {Park}, {Pasquato}, {Peltzer}, {Peralta}, {P{\'e}turaud}, {Pieniluoma}, {Pigozzi}, {Poels}, {Prat}, {Prod'homme}, {Raison}, {Rebordao}, {Risquez}, {Rocca-Volmerange}, {Rosen},
  {Ruiz-Fuertes}, {Russo}, {Sembay}, {Serraller Vizcaino}, {Short}, {Siebert}, {Silva}, {Sinachopoulos}, {Slezak}, {Soffel}, {Sosnowska}, {Strai{\v{z}}ys}, {ter Linden}, {Terrell}, {Theil}, {Tiede}, {Troisi}, {Tsalmantza}, {Tur}, {Vaccari}, {Vachier}, {Valles}, {Van Hamme}, {Veltz}, {Virtanen}, {Wallut}, {Wichmann}, {Wilkinson}, {Ziaeepour}, \& {Zschocke}}]{2016A&A...595A...1G}
{Gaia Collaboration}, {Prusti}, T., {de Bruijne}, J.~H.~J., {et~al.} 2016, \href{http://dx.doi.org/10.1051/0004-6361/201629272}{\color{magenta}\aap}, \href{https://ui.adsabs.harvard.edu/abs/2016A&A...595A...1G}{595, A1}

\bibitem[{{Gavil{\'a}n} {et~al.}(2005){Gavil{\'a}n}, {Buell}, \& {Moll{\'a}}}]{2005A&A...432..861G}
{Gavil{\'a}n}, M., {Buell}, J.~F., \& {Moll{\'a}}, M. 2005, \href{http://dx.doi.org/10.1051/0004-6361:20041949}{\color{magenta}\aap}, \href{https://ui.adsabs.harvard.edu/abs/2005A&A...432..861G}{432, 861}

\bibitem[{{Georgy} {et~al.}(2013){Georgy}, {Ekstr{\"o}m}, {Eggenberger}, {Meynet}, {Haemmerl{\'e}}, {Maeder}, {Granada}, {Groh}, {Hirschi}, {Mowlavi}, {Yusof}, {Charbonnel}, {Decressin}, \& {Barblan}}]{Georgy2012}
{Georgy}, C., {Ekstr{\"o}m}, S., {Eggenberger}, P., {et~al.} 2013, \href{http://dx.doi.org/10.1051/0004-6361/201322178}{\color{magenta}\aap}, \href{https://ui.adsabs.harvard.edu/abs/2013A&A...558A.103G}{558, A103}

\bibitem[{{Gratton} {et~al.}(2000){Gratton}, {Sneden}, {Carretta}, \& {Bragaglia}}]{2000A&A...354..169G}
{Gratton}, R.~G., {Sneden}, C., {Carretta}, E., \& {Bragaglia}, A. 2000, \aap, \href{https://ui.adsabs.harvard.edu/abs/2000A&A...354..169G}{354, 169}

\bibitem[{{Groh} {et~al.}(2019){Groh}, {Ekstr{\"o}m}, {Georgy}, {Meynet}, {Choplin}, {Eggenberger}, {Hirschi}, {Maeder}, {Murphy}, {Boian}, \& {Farrell}}]{Groh2019}
{Groh}, J.~H., {Ekstr{\"o}m}, S., {Georgy}, C., {et~al.} 2019, \href{http://dx.doi.org/10.1051/0004-6361/201833720}{\color{magenta}\aap}, \href{https://ui.adsabs.harvard.edu/abs/2019A&A...627A..24G}{627, A24}

\bibitem[{{Hansen} {et~al.}(2019){Hansen}, {Hansen}, {Koch}, {Beers}, {Nordstr{\"o}m}, {Placco}, \& {Andersen}}]{Hansen2019}
{Hansen}, C.~J., {Hansen}, T.~T., {Koch}, A., {et~al.} 2019, \href{http://dx.doi.org/10.1051/0004-6361/201834601}{\color{magenta}\aap}, \href{https://ui.adsabs.harvard.edu/abs/2019A&A...623A.128H}{623, A128}

\bibitem[{{Hirschi}(2007)}]{2007A&A...461..571H}
{Hirschi}, R. 2007, \href{http://dx.doi.org/10.1051/0004-6361:20065356}{\color{magenta}\aap}, \href{https://ui.adsabs.harvard.edu/abs/2007A&A...461..571H}{461, 571}

\bibitem[{{Hirschi} {et~al.}(2005){Hirschi}, {Meynet}, \& {Maeder}}]{2005A&A...433.1013H}
{Hirschi}, R., {Meynet}, G., \& {Maeder}, A. 2005, \href{http://dx.doi.org/10.1051/0004-6361:20041554}{\color{magenta}\aap}, \href{https://ui.adsabs.harvard.edu/abs/2005A&A...433.1013H}{433, 1013}

\bibitem[{{Hoyle}(1954)}]{1954ApJS....1..121H}
{Hoyle}, F. 1954, \href{http://dx.doi.org/10.1086/190005}{\color{magenta}\apjs}, \href{https://ui.adsabs.harvard.edu/abs/1954ApJS....1..121H}{1, 121}

\bibitem[{Huber \& Herzberg(1979)}]{Huber1979}
Huber, K.~P. \& Herzberg, G. 1979, Constants of diatomic molecules (Boston, MA: Springer US), 8--689

\bibitem[{{Iben} \& {Renzini}(1984)}]{1984PhR...105..329I}
{Iben}, I. \& {Renzini}, A. 1984, \href{http://dx.doi.org/10.1016/0370-1573(84)90142-X}{\color{magenta}\physrep}, \href{https://ui.adsabs.harvard.edu/abs/1984PhR...105..329I}{105, 329}

\bibitem[{{Iben}(1964)}]{1964ApJ...140.1631I}
{Iben}, Icko, J. 1964, \href{http://dx.doi.org/10.1086/148077}{\color{magenta}\apj}, \href{https://ui.adsabs.harvard.edu/abs/1964ApJ...140.1631I}{140, 1631}

\bibitem[{{Ishigaki} {et~al.}(2014){Ishigaki}, {Aoki}, {Arimoto}, \& {Okamoto}}]{2014A&A...562A.146I}
{Ishigaki}, M.~N., {Aoki}, W., {Arimoto}, N., \& {Okamoto}, S. 2014, \href{http://dx.doi.org/10.1051/0004-6361/201322796}{\color{magenta}\aap}, \href{https://ui.adsabs.harvard.edu/abs/2014A&A...562A.146I}{562, A146}

\bibitem[{{Israelian} {et~al.}(2004){Israelian}, {Ecuvillon}, {Rebolo}, {Garc{\'\i}a-L{\'o}pez}, {Bonifacio}, \& {Molaro}}]{2004A&A...421..649I}
{Israelian}, G., {Ecuvillon}, A., {Rebolo}, R., {et~al.} 2004, \href{http://dx.doi.org/10.1051/0004-6361:20047132}{\color{magenta}\aap}, \href{https://ui.adsabs.harvard.edu/abs/2004A&A...421..649I}{421, 649}

\bibitem[{{Karakas} \& {Lattanzio}(2014)}]{2014PASA...31...30K}
{Karakas}, A.~I. \& {Lattanzio}, J.~C. 2014, \href{http://dx.doi.org/10.1017/pasa.2014.21}{\color{magenta}\pasa}, \href{https://ui.adsabs.harvard.edu/abs/2014PASA...31...30K}{31, e030}

\bibitem[{{Kennicutt}(1998)}]{1998ApJ...498..541K}
{Kennicutt}, Robert~C., J. 1998, \href{http://dx.doi.org/10.1086/305588}{\color{magenta}\apj}, \href{https://ui.adsabs.harvard.edu/abs/1998ApJ...498..541K}{498, 541}

\bibitem[{{Kobayashi} {et~al.}(2020){Kobayashi}, {Karakas}, \& {Lugaro}}]{2020ApJ...900..179K}
{Kobayashi}, C., {Karakas}, A.~I., \& {Lugaro}, M. 2020, \href{http://dx.doi.org/10.3847/1538-4357/abae65}{\color{magenta}\apj}, \href{https://ui.adsabs.harvard.edu/abs/2020ApJ...900..179K}{900, 179}

\bibitem[{{Kobayashi} {et~al.}(2006){Kobayashi}, {Umeda}, {Nomoto}, {Tominaga}, \& {Ohkubo}}]{2006ApJ...653.1145K}
{Kobayashi}, C., {Umeda}, H., {Nomoto}, K., {Tominaga}, N., \& {Ohkubo}, T. 2006, \href{http://dx.doi.org/10.1086/508914}{\color{magenta}\apj}, \href{https://ui.adsabs.harvard.edu/abs/2006ApJ...653.1145K}{653, 1145}

\bibitem[{{Koch-Hansen} {et~al.}(2021){Koch-Hansen}, {Hansen}, {Lombardo}, {Bonifacio}, {Hanke}, \& {Caffau}}]{2021A&A...645A..64K}
{Koch-Hansen}, A.~J., {Hansen}, C.~J., {Lombardo}, L., {et~al.} 2021, \href{http://dx.doi.org/10.1051/0004-6361/202039711}{\color{magenta}\aap}, \href{https://ui.adsabs.harvard.edu/abs/2021A&A...645A..64K}{645, A64}

\bibitem[{{Korn} {et~al.}(2006){Korn}, {Grundahl}, {Richard}, {Barklem}, {Mashonkina}, {Collet}, {Piskunov}, \& {Gustafsson}}]{Korn2006Natur}
{Korn}, A.~J., {Grundahl}, F., {Richard}, O., {et~al.} 2006, \href{http://dx.doi.org/10.1038/nature05011}{\color{magenta}\nat}, \href{https://ui.adsabs.harvard.edu/abs/2006Natur.442..657K}{442, 657}

\bibitem[{{Korn} {et~al.}(2007){Korn}, {Grundahl}, {Richard}, {Mashonkina}, {Barklem}, {Collet}, {Gustafsson}, \& {Piskunov}}]{Korn2007}
{Korn}, A.~J., {Grundahl}, F., {Richard}, O., {et~al.} 2007, \href{http://dx.doi.org/10.1086/523098}{\color{magenta}\apj}, \href{https://ui.adsabs.harvard.edu/abs/2007ApJ...671..402K}{671, 402}

\bibitem[{{Kroupa} {et~al.}(1993){Kroupa}, {Tout}, \& {Gilmore}}]{1993MNRAS.262..545K}
{Kroupa}, P., {Tout}, C.~A., \& {Gilmore}, G. 1993, \href{http://dx.doi.org/10.1093/mnras/262.3.545}{\color{magenta}\mnras}, \href{https://ui.adsabs.harvard.edu/abs/1993MNRAS.262..545K}{262, 545}

\bibitem[{{Kurucz}(2005)}]{2005MSAIS...8...14K}
{Kurucz}, R.~L. 2005, Memorie della Societa Astronomica Italiana Supplementi, \href{https://ui.adsabs.harvard.edu/abs/2005MSAIS...8...14K}{8, 14}

\bibitem[{{Lagarde} {et~al.}(2019){Lagarde}, {Reyl{\'e}}, {Robin}, {Tautvai{\v{s}}ien{\.{e}}}, {Drazdauskas}, {Mikolaitis}, {Minkevi{\v{c}}i{\={u}}t{\.{e}}}, {Stonkut{\.{e}}}, {Chorniy}, {Bagdonas}, {Miglio}, {Nasello}, {Gilmore}, {Randich}, {Bensby}, {Bragaglia}, {Flaccomio}, {Francois}, {Korn}, {Pancino}, {Smiljanic}, {Bayo}, {Carraro}, {Costado}, {Jim{\'e}nez-Esteban}, {Jofr{\'e}}, {Martell}, {Masseron}, {Monaco}, {Morbidelli}, {Sbordone}, {Sousa}, \& {Zaggia}}]{2019A&A...621A..24L}
{Lagarde}, N., {Reyl{\'e}}, C., {Robin}, A.~C., {et~al.} 2019, \href{http://dx.doi.org/10.1051/0004-6361/201732433}{\color{magenta}\aap}, \href{https://ui.adsabs.harvard.edu/abs/2019A&A...621A..24L}{621, A24}

\bibitem[{{Lagarde} {et~al.}(2017){Lagarde}, {Robin}, {Reyl{\'e}}, \& {Nasello}}]{2017A&A...601A..27L}
{Lagarde}, N., {Robin}, A.~C., {Reyl{\'e}}, C., \& {Nasello}, G. 2017, \href{http://dx.doi.org/10.1051/0004-6361/201630253}{\color{magenta}\aap}, \href{https://ui.adsabs.harvard.edu/abs/2017A&A...601A..27L}{601, A27}

\bibitem[{{Limongi} \& {Chieffi}(2018)}]{2018ApJS..237...13L}
{Limongi}, M. \& {Chieffi}, A. 2018, \href{http://dx.doi.org/10.3847/1538-4365/aacb24}{\color{magenta}\apjs}, \href{https://ui.adsabs.harvard.edu/abs/2018ApJS..237...13L}{237, 13}

\bibitem[{{Lodders} {et~al.}(2009){Lodders}, {Palme}, \& {Gail}}]{2009LanB...4B..712L}
{Lodders}, K., {Palme}, H., \& {Gail}, H.~P. 2009, \href{http://dx.doi.org/10.1007/978-3-540-88055-4_34}{\color{magenta}Landolt B\&ouml;rnstein}, \href{https://ui.adsabs.harvard.edu/abs/2009LanB...4B..712L}{4B, 712}

\bibitem[{{Lombardo} {et~al.}(2022){Lombardo}, {Bonifacio}, {Fran{\c{c}}ois}, {Hansen}, {Caffau}, {Hanke}, {Sk{\'u}lad{\'o}ttir}, {Arcones}, {Eichler}, {Reichert}, {Psaltis}, {Koch Hansen}, \& {Sbordone}}]{2022A&A...665A..10L}
{Lombardo}, L., {Bonifacio}, P., {Fran{\c{c}}ois}, P., {et~al.} 2022, \href{http://dx.doi.org/10.1051/0004-6361/202243932}{\color{magenta}\aap}, \href{https://ui.adsabs.harvard.edu/abs/2022A&A...665A..10L}{665, A10}

\bibitem[{{Maeder}(1992)}]{Maeder1992}
{Maeder}, A. 1992, \aap, \href{https://ui.adsabs.harvard.edu/abs/1992A&A...264..105M}{264, 105}

\bibitem[{{Maeder} \& {Meynet}(2000)}]{MaederMeynet2000}
{Maeder}, A. \& {Meynet}, G. 2000, \href{http://dx.doi.org/10.48550/arXiv.astro-ph/0006405}{\color{magenta}\aap}, \href{https://ui.adsabs.harvard.edu/abs/2000A&A...361..159M}{361, 159}

\bibitem[{{Maeder} \& {Meynet}(2001)}]{2001A&A...373..555M}
{Maeder}, A. \& {Meynet}, G. 2001, \href{http://dx.doi.org/10.1051/0004-6361:20010596}{\color{magenta}\aap}, \href{https://ui.adsabs.harvard.edu/abs/2001A&A...373..555M}{373, 555}

\bibitem[{{Martell} {et~al.}(2008){Martell}, {Smith}, \& {Briley}}]{2008AJ....136.2522M}
{Martell}, S.~L., {Smith}, G.~H., \& {Briley}, M.~M. 2008, \href{http://dx.doi.org/10.1088/0004-6256/136/6/2522}{\color{magenta}\aj}, \href{https://ui.adsabs.harvard.edu/abs/2008AJ....136.2522M}{136, 2522}

\bibitem[{{Mashonkina} {et~al.}(2017){Mashonkina}, {Jablonka}, {Pakhomov}, {Sitnova}, \& {North}}]{Mashonkina2017A&A...604A.129M}
{Mashonkina}, L., {Jablonka}, P., {Pakhomov}, Y., {Sitnova}, T., \& {North}, P. 2017, \href{http://dx.doi.org/10.1051/0004-6361/201730779}{\color{magenta}\aap}, \href{https://ui.adsabs.harvard.edu/abs/2017A&A...604A.129M}{604, A129}

\bibitem[{{Masseron} {et~al.}(2014){Masseron}, {Plez}, {Van Eck}, {Colin}, {Daoutidis}, {Godefroid}, {Coheur}, {Bernath}, {Jorissen}, \& {Christlieb}}]{Masseron2014A&A...571A..47M}
{Masseron}, T., {Plez}, B., {Van Eck}, S., {et~al.} 2014, \href{http://dx.doi.org/10.1051/0004-6361/201423956}{\color{magenta}\aap}, \href{https://ui.adsabs.harvard.edu/abs/2014A&A...571A..47M}{571, A47}

\bibitem[{{McWilliam} {et~al.}(1995){McWilliam}, {Preston}, {Sneden}, \& {Searle}}]{1995AJ....109.2757M}
{McWilliam}, A., {Preston}, G.~W., {Sneden}, C., \& {Searle}, L. 1995, \href{http://dx.doi.org/10.1086/117486}{\color{magenta}\aj}, \href{https://ui.adsabs.harvard.edu/abs/1995AJ....109.2757M}{109, 2757}

\bibitem[{{Meynet} {et~al.}(2006){Meynet}, {Ekstr{\"o}m}, \& {Maeder}}]{2006A&A...447..623M}
{Meynet}, G., {Ekstr{\"o}m}, S., \& {Maeder}, A. 2006, \href{http://dx.doi.org/10.1051/0004-6361:20053070}{\color{magenta}\aap}, \href{https://ui.adsabs.harvard.edu/abs/2006A&A...447..623M}{447, 623}

\bibitem[{{Meynet} \& {Maeder}(2002)}]{2002A&A...390..561M}
{Meynet}, G. \& {Maeder}, A. 2002, \href{http://dx.doi.org/10.1051/0004-6361:20020755}{\color{magenta}\aap}, \href{https://ui.adsabs.harvard.edu/abs/2002A&A...390..561M}{390, 561}

\bibitem[{{Mucciarelli} {et~al.}(2022){Mucciarelli}, {Monaco}, {Bonifacio}, {Salaris}, {Deal}, {Spite}, {Richard}, \& {Lallement}}]{2022A&A...661A.153M}
{Mucciarelli}, A., {Monaco}, L., {Bonifacio}, P., {et~al.} 2022, \href{http://dx.doi.org/10.1051/0004-6361/202142889}{\color{magenta}\aap}, \href{https://ui.adsabs.harvard.edu/abs/2022A&A...661A.153M}{661, A153}

\bibitem[{{Nissen} {et~al.}(2014){Nissen}, {Chen}, {Carigi}, {Schuster}, \& {Zhao}}]{2014A&A...568A..25N}
{Nissen}, P.~E., {Chen}, Y.~Q., {Carigi}, L., {Schuster}, W.~J., \& {Zhao}, G. 2014, \href{http://dx.doi.org/10.1051/0004-6361/201424184}{\color{magenta}\aap}, \href{https://ui.adsabs.harvard.edu/abs/2014A&A...568A..25N}{568, A25}

\bibitem[{{Nissen} \& {Gustafsson}(2018)}]{2018A&ARv..26....6N}
{Nissen}, P.~E. \& {Gustafsson}, B. 2018, \href{http://dx.doi.org/10.1007/s00159-018-0111-3}{\color{magenta}\aapr}, \href{https://ui.adsabs.harvard.edu/abs/2018A&ARv..26....6N}{26, 6}

\bibitem[{{Nissen} {et~al.}(2002){Nissen}, {Primas}, {Asplund}, \& {Lambert}}]{Nissen2002A&A...390..235N}
{Nissen}, P.~E., {Primas}, F., {Asplund}, M., \& {Lambert}, D.~L. 2002, \href{http://dx.doi.org/10.1051/0004-6361:20020736}{\color{magenta}\aap}, \href{https://ui.adsabs.harvard.edu/abs/2002A&A...390..235N}{390, 235}

\bibitem[{{Nomoto} {et~al.}(2013){Nomoto}, {Kobayashi}, \& {Tominaga}}]{2013ARA&A..51..457N}
{Nomoto}, K., {Kobayashi}, C., \& {Tominaga}, N. 2013, \href{http://dx.doi.org/10.1146/annurev-astro-082812-140956}{\color{magenta}\araa}, \href{https://ui.adsabs.harvard.edu/abs/2013ARA&A..51..457N}{51, 457}

\bibitem[{{Papadopoulos}(2010)}]{2010ApJ...720..226P}
{Papadopoulos}, P.~P. 2010, \href{http://dx.doi.org/10.1088/0004-637X/720/1/226}{\color{magenta}\apj}, \href{https://ui.adsabs.harvard.edu/abs/2010ApJ...720..226P}{720, 226}

\bibitem[{{Patton} \& {Sukhbold}(2020)}]{Patton2020}
{Patton}, R.~A. \& {Sukhbold}, T. 2020, \href{http://dx.doi.org/10.1093/mnras/staa3029}{\color{magenta}\mnras}, \href{https://ui.adsabs.harvard.edu/abs/2020MNRAS.499.2803P}{499, 2803}

\bibitem[{{Placco} {et~al.}(2021){Placco}, {Sneden}, {Roederer}, {Lawler}, {Den Hartog}, {Hejazi}, {Maas}, \& {Bernath}}]{2021RNAAS...5...92P}
{Placco}, V.~M., {Sneden}, C., {Roederer}, I.~U., {et~al.} 2021, \href{http://dx.doi.org/10.3847/2515-5172/abf651}{\color{magenta}Research Notes of the American Astronomical Society}, \href{https://ui.adsabs.harvard.edu/abs/2021RNAAS...5...92P}{5, 92}

\bibitem[{{Popa} {et~al.}(2023){Popa}, {Hoppe}, {Bergemann}, {Hansen}, {Plez}, \& {Beers}}]{2023A&A...670A..25P}
{Popa}, S.~A., {Hoppe}, R., {Bergemann}, M., {et~al.} 2023, \href{http://dx.doi.org/10.1051/0004-6361/202245503}{\color{magenta}\aap}, \href{https://ui.adsabs.harvard.edu/abs/2023A&A...670A..25P}{670, A25}

\bibitem[{{Prantzos} {et~al.}(2018){Prantzos}, {Abia}, {Limongi}, {Chieffi}, \& {Cristallo}}]{2018MNRAS.476.3432P}
{Prantzos}, N., {Abia}, C., {Limongi}, M., {Chieffi}, A., \& {Cristallo}, S. 2018, \href{http://dx.doi.org/10.1093/mnras/sty316}{\color{magenta}\mnras}, \href{https://ui.adsabs.harvard.edu/abs/2018MNRAS.476.3432P}{476, 3432}

\bibitem[{{Renzini} \& {Voli}(1981)}]{1981A&A....94..175R}
{Renzini}, A. \& {Voli}, M. 1981, \aap, \href{https://ui.adsabs.harvard.edu/abs/1981A&A....94..175R}{94, 175}

\bibitem[{{Roederer} {et~al.}(2014){Roederer}, {Preston}, {Thompson}, {Shectman}, {Sneden}, {Burley}, \& {Kelson}}]{2014AJ....147..136R}
{Roederer}, I.~U., {Preston}, G.~W., {Thompson}, I.~B., {et~al.} 2014, \href{http://dx.doi.org/10.1088/0004-6256/147/6/136}{\color{magenta}\aj}, \href{https://ui.adsabs.harvard.edu/abs/2014AJ....147..136R}{147, 136}

\bibitem[{{Romano}(2022)}]{2022A&ARv..30....7R}
{Romano}, D. 2022, \href{http://dx.doi.org/10.1007/s00159-022-00144-z}{\color{magenta}\aapr}, \href{https://ui.adsabs.harvard.edu/abs/2022A&ARv..30....7R}{30, 7}

\bibitem[{{Romano} {et~al.}(2020){Romano}, {Franchini}, {Grisoni}, {Spitoni}, {Matteucci}, \& {Morossi}}]{2020A&A...639A..37R}
{Romano}, D., {Franchini}, M., {Grisoni}, V., {et~al.} 2020, \href{http://dx.doi.org/10.1051/0004-6361/202037972}{\color{magenta}\aap}, \href{https://ui.adsabs.harvard.edu/abs/2020A&A...639A..37R}{639, A37}

\bibitem[{{Romano} {et~al.}(2010){Romano}, {Karakas}, {Tosi}, \& {Matteucci}}]{2010A&A...522A..32R}
{Romano}, D., {Karakas}, A.~I., {Tosi}, M., \& {Matteucci}, F. 2010, \href{http://dx.doi.org/10.1051/0004-6361/201014483}{\color{magenta}\aap}, \href{https://ui.adsabs.harvard.edu/abs/2010A&A...522A..32R}{522, A32}

\bibitem[{{Romano} {et~al.}(2019){Romano}, {Matteucci}, {Zhang}, {Ivison}, \& {Ventura}}]{2019MNRAS.490.2838R}
{Romano}, D., {Matteucci}, F., {Zhang}, Z.-Y., {Ivison}, R.~J., \& {Ventura}, P. 2019, \href{http://dx.doi.org/10.1093/mnras/stz2741}{\color{magenta}\mnras}, \href{https://ui.adsabs.harvard.edu/abs/2019MNRAS.490.2838R}{490, 2838}

\bibitem[{{Romano} {et~al.}(2017){Romano}, {Matteucci}, {Zhang}, {Papadopoulos}, \& {Ivison}}]{2017MNRAS.470..401R}
{Romano}, D., {Matteucci}, F., {Zhang}, Z.~Y., {Papadopoulos}, P.~P., \& {Ivison}, R.~J. 2017, \href{http://dx.doi.org/10.1093/mnras/stx1197}{\color{magenta}\mnras}, \href{https://ui.adsabs.harvard.edu/abs/2017MNRAS.470..401R}{470, 401}

\bibitem[{{Roy} {et~al.}(2021){Roy}, {Dopita}, {Krumholz}, {Kewley}, {Sutherland}, \& {Heger}}]{2021MNRAS.502.4359R}
{Roy}, A., {Dopita}, M.~A., {Krumholz}, M.~R., {et~al.} 2021, \href{http://dx.doi.org/10.1093/mnras/stab376}{\color{magenta}\mnras}, \href{https://ui.adsabs.harvard.edu/abs/2021MNRAS.502.4359R}{502, 4359}

\bibitem[{{Sbordone} {et~al.}(2014){Sbordone}, {Caffau}, {Bonifacio}, \& {Duffau}}]{2014A&A...564A.109S}
{Sbordone}, L., {Caffau}, E., {Bonifacio}, P., \& {Duffau}, S. 2014, \href{http://dx.doi.org/10.1051/0004-6361/201322430}{\color{magenta}\aap}, \href{https://ui.adsabs.harvard.edu/abs/2014A&A...564A.109S}{564, A109}

\bibitem[{{Schlafly} \& {Finkbeiner}(2011)}]{2011ApJ...737..103S}
{Schlafly}, E.~F. \& {Finkbeiner}, D.~P. 2011, \href{http://dx.doi.org/10.1088/0004-637X/737/2/103}{\color{magenta}\apj}, \href{https://ui.adsabs.harvard.edu/abs/2011ApJ...737..103S}{737, 103}

\bibitem[{{Shejeelammal} \& {Goswami}(2024)}]{2024MNRAS.527.2323S}
{Shejeelammal}, J. \& {Goswami}, A. 2024, \href{http://dx.doi.org/10.1093/mnras/stad3290}{\color{magenta}\mnras}, \href{https://ui.adsabs.harvard.edu/abs/2024MNRAS.527.2323S}{527, 2323}

\bibitem[{{Sibony} {et~al.}(2024){Sibony}, {Shepherd}, {Yusof}, {Hirschi}, {Chambers}, {Tsiatsiou}, {Nandal}, {Sciarini}, {Moyano}, {B{\'e}trisey}, {Buldgen}, {Georgy}, {Ekstr{\"o}m}, {Eggenberger}, \& {Meynet}}]{Sibony2024}
{Sibony}, Y., {Shepherd}, K., {Yusof}, N., {et~al.} 2024, \aap, submitted

\bibitem[{{Sneden}(1973)}]{1973PhDT.......180S}
{Sneden}, C.~A. 1973, \href{https://ui.adsabs.harvard.edu/abs/1973PhDT.......180S}{{Carbon and Nitrogen Abundances in Metal-Poor Stars.}}, PhD thesis, University of Texas, Austin

\bibitem[{{Spite} {et~al.}(2005){Spite}, {Cayrel}, {Plez}, {Hill}, {Spite}, {Depagne}, {Fran{\c{c}}ois}, {Bonifacio}, {Barbuy}, {Beers}, {Andersen}, {Molaro}, {Nordstr{\"o}m}, \& {Primas}}]{2005A&A...430..655S}
{Spite}, M., {Cayrel}, R., {Plez}, B., {et~al.} 2005, \href{http://dx.doi.org/10.1051/0004-6361:20041274}{\color{magenta}\aap}, \href{https://ui.adsabs.harvard.edu/abs/2005A&A...430..655S}{430, 655}

\bibitem[{{Spitoni} {et~al.}(2019){Spitoni}, {Silva Aguirre}, {Matteucci}, {Calura}, \& {Grisoni}}]{2019A&A...623A..60S}
{Spitoni}, E., {Silva Aguirre}, V., {Matteucci}, F., {Calura}, F., \& {Grisoni}, V. 2019, \href{http://dx.doi.org/10.1051/0004-6361/201834188}{\color{magenta}\aap}, \href{https://ui.adsabs.harvard.edu/abs/2019A&A...623A..60S}{623, A60}

\bibitem[{{Spitoni} {et~al.}(2021){Spitoni}, {Verma}, {Silva Aguirre}, {Vincenzo}, {Matteucci}, {Vai{\v{c}}ekauskait{\.{e}}}, {Palla}, {Grisoni}, \& {Calura}}]{2021A&A...647A..73S}
{Spitoni}, E., {Verma}, K., {Silva Aguirre}, V., {et~al.} 2021, \href{http://dx.doi.org/10.1051/0004-6361/202039864}{\color{magenta}\aap}, \href{https://ui.adsabs.harvard.edu/abs/2021A&A...647A..73S}{647, A73}

\bibitem[{{Storey} \& {Zeippen}(2000)}]{2000MNRAS.312..813S}
{Storey}, P.~J. \& {Zeippen}, C.~J. 2000, \href{http://dx.doi.org/10.1046/j.1365-8711.2000.03184.x}{\color{magenta}\mnras}, \href{https://ui.adsabs.harvard.edu/abs/2000MNRAS.312..813S}{312, 813}

\bibitem[{{Tsiatsiou} {et~al.}(2024){Tsiatsiou}, {Sibony}, {Nandal}, {Sciarini}, {Hirai}, {Ekstrom}, {Farrell}, {Murphy}, {Choplin}, {Hirschi}, {Chiappini}, {Liu}, {Bromm}, {Groh}, \& {Meynet}}]{2024arXiv240416512T}
{Tsiatsiou}, S., {Sibony}, Y., {Nandal}, D., {et~al.} 2024, \href{https://ui.adsabs.harvard.edu/abs/2024arXiv240416512T}{\href{http://dx.doi.org/10.48550/arXiv.2404.16512}{\color{magenta}arXiv e-prints}, arXiv:2404.16512}

\bibitem[{{Umeda} \& {Nomoto}(2002)}]{2002ApJ...565..385U}
{Umeda}, H. \& {Nomoto}, K. 2002, \href{http://dx.doi.org/10.1086/323946}{\color{magenta}\apj}, \href{https://ui.adsabs.harvard.edu/abs/2002ApJ...565..385U}{565, 385}

\bibitem[{{Vincenzo} {et~al.}(2016){Vincenzo}, {Belfiore}, {Maiolino}, {Matteucci}, \& {Ventura}}]{2016MNRAS.458.3466V}
{Vincenzo}, F., {Belfiore}, F., {Maiolino}, R., {Matteucci}, F., \& {Ventura}, P. 2016, \href{http://dx.doi.org/10.1093/mnras/stw532}{\color{magenta}\mnras}, \href{https://ui.adsabs.harvard.edu/abs/2016MNRAS.458.3466V}{458, 3466}

\bibitem[{{Wang} {et~al.}(2021){Wang}, {Nordlander}, {Asplund}, {Amarsi}, {Lind}, \& {Zhou}}]{2021MNRAS.500.2159W}
{Wang}, E.~X., {Nordlander}, T., {Asplund}, M., {et~al.} 2021, \href{http://dx.doi.org/10.1093/mnras/staa3381}{\color{magenta}\mnras}, \href{https://ui.adsabs.harvard.edu/abs/2021MNRAS.500.2159W}{500, 2159}

\bibitem[{{Woosley} {et~al.}(2002){Woosley}, {Heger}, \& {Weaver}}]{2002RvMP...74.1015W}
{Woosley}, S.~E., {Heger}, A., \& {Weaver}, T.~A. 2002, \href{http://dx.doi.org/10.1103/RevModPhys.74.1015}{\color{magenta}Reviews of Modern Physics}, \href{https://ui.adsabs.harvard.edu/abs/2002RvMP...74.1015W}{74, 1015}

\end{thebibliography}
%\newpage
%\include{appendix}

\begin{appendix}
\onecolumn

\section{Abundances}

\begin{table*}[h!]
    \centering
    \caption{Chemical abundances derived in this work.}
\begin{tabular}{lcccccccc}
\hline\hline
Star         &  [Fe/H]& [\ion{Fe}{ii}/H] & A(C) & [C/Fe] & A(N)  & [N/Fe]&  A(O) & [O/Fe] \\
%& (dex) & (dex) & (dex) & (dex)& (dex) & (dex) & (dex) &\\
\hline                                        
CES\,0031$-$1647 &  $-$2.49 & $-$2.31    & 5.56 & $-$0.34  &  5.89 &  0.52   &  7.00 &  0.76  \\
CES\,0045$-$0932 &  $-$2.95 & $-$2.80    &   -- &    --    &  4.04 & $-$0.87 &    -- &   --   \\
CES\,0048$-$1041 &  $-$2.48 & $-$2.33    & 5.54 & $-$0.37  &  5.87 &  0.49   &  7.05 &  0.80  \\
CES\,0055$-$3345 &  $-$2.36 & $-$2.24    & 6.27 &  0.24    &  4.80 & $-$0.70 &    -- &   --   \\
CES\,0059$-$4524 &  $-$2.39 & $-$2.26    & 6.37 &  0.37    &    -- &    --   &     -- &   --   \\
CES\,0102$-$6143 &  $-$2.86 & $-$2.84    & 6.02 &  0.49    &    -- &    --   &    -- &   --   \\
CES\,0107$-$6125 &  $-$2.59 & $-$2.56    & 6.22 &  0.42    &    -- &    --   &    -- &   --   \\
CES\,0109$-$0443 &  $-$3.23 & $-$3.24    & 5.67 &  0.51    &  4.78 &  0.15   &    -- &   --   \\
CES\,0215$-$2554 &  $-$2.73 & $-$2.54    & 5.17 & $-$0.49  &  5.62 &  0.49   &    -- &   --   \\
CES\,0221$-$2130 &  $-$1.99 & $-$1.79    & 6.05 & $-$0.35  &  6.04 &  0.17   &    -- &   --   \\
CES\,0242$-$0754 &  $-$2.90 & $-$2.85    & 5.57 &  0.08    &  5.95 &  0.99   &    -- &   --   \\
CES\,0301$+$0616 &  $-$2.93 & $-$2.85    & 5.97 &  0.51    &  4.50 & $-$0.43 &    -- &   --   \\
CES\,0338$-$2402 &  $-$2.81 & $-$2.72    &   -- &    --    &  4.42 & $-$0.63 &    -- &   --   \\
CES\,0413$+$0636 &  $-$2.24 & $-$2.21    & 6.25 &  0.10    &  6.91 &  1.29   &  7.27 &  0.78  \\
CES\,0419$-$3651 &  $-$2.81 & $-$2.7     & 6.06 &  0.48    &  5.33 &  0.28   &    -- &   --   \\
CES\,0422$-$3715 &  $-$2.45 & $-$2.37    & 6.26 &  0.32    &  4.82 & $-$0.59 &  7.02 &  0.74  \\
CES\,0424$-$1501 &  $-$1.79 & $-$1.64    & 6.45 & $-$0.15  &  5.96 & $-$0.11 &    -- &   --   \\
CES\,0430$-$1334 &  $-$2.09 & $-$2.13    & 6.65 &  0.35    &  5.72 & $-$0.05 &    -- &   --   \\
CES\,0444$-$1228 &  $-$2.54 & $-$2.35    & 5.42 & $-$0.43  &  6.02 &  0.70   &  6.98 &  0.79  \\
CES\,0518$-$3817 &  $-$2.49 & $-$2.47    & 6.35 &  0.45    &    -- &    --   &    -- &   --   \\
CES\,0527$-$2052 &  $-$2.75 & $-$2.53    & 5.63 & $-$0.01  &  5.63 &  0.52   &    -- &   --   \\
CES\,0547$-$1739 &  $-$2.05 & $-$1.81    & 5.78 & $-$0.56  &  6.43 &  0.62   &  7.34 &  0.66  \\
CES\,0747$-$0405 &  $-$2.25 & $-$1.98    & 5.68 & $-$0.46  &  6.13 &  0.52   &  7.19 &  0.71  \\
CES\,0900$-$6222 &  $-$2.11 & $-$1.89    & 5.70 & $-$0.58  &  6.44 &  0.69   &  7.40 &  0.78  \\
CES\,0908$-$6607 &  $-$2.62 & $-$2.54    & 5.58 & $-$0.19  &  6.93 &  1.69   &  6.84 &  0.73  \\
CES\,0919$-$6958 &  $-$2.46 & $-$2.42    & 5.82 & $-$0.11  &  6.86 &  1.46   &  6.90 &  0.63  \\
CES\,1116$-$7250 &  $-$2.74 & $-$2.39    & 5.43 & $-$0.22  &  5.94 &  0.82   &  6.86 &  0.87  \\
CES\,1221$-$0328 &  $-$2.96 & $-$2.74    &   -- &    --    &  4.63 & $-$0.27 &    -- &   --   \\
CES\,1222$+$1136 &  $-$2.91 & $-$2.82    & 5.67 &  0.19    &  5.48 &  0.53   &    -- &   --   \\
CES\,1226$+$0518 &  $-$2.38 & $-$2.33    &   -- &    --    &  4.95 & $-$0.53 &    -- &   --   \\
CES\,1228$+$1220 &  $-$2.32 & $-$2.11    & 5.82 & $-$0.25  &  5.91 &  0.37   &  7.23 &  0.82  \\
CES\,1237$+$1922 &  $-$3.19 & $-$3.04    & 5.05 & $-$0.15  &  5.30 &  0.63   &    -- &   --   \\
CES\,1245$-$2425 &  $-$2.85 & $-$2.72    & 5.85 &  0.31    &  4.35 & $-$0.66 &    -- &   --   \\
CES\,1322$-$1355 &  $-$2.93 & $-$2.78    & 5.26 & $-$0.20  &  5.64 &  0.71   &  6.63 &  0.83  \\
CES\,1402$+$0941 &  $-$2.79 & $-$2.72    &   -- &    --    &    -- &    --   &  6.79 &  0.85  \\
CES\,1405$-$1451 &  $-$1.87 & $-$1.71    &   -- &    --    &  6.05 &  0.06   &  7.49 &  0.63  \\
CES\,1413$-$7609 &  $-$2.52 & $-$2.42    & 5.84 & $-$0.03  &  5.70 &  0.36   &  7.01 &  0.80  \\
CES\,1427$-$2214 &  $-$3.05 & $-$2.94    &   -- &    --    &  4.12 & $-$0.69 &  6.69 &  1.01  \\
CES\,1436$-$2906 &  $-$2.15 & $-$2.23    &   -- &    --    &  4.65 & $-$1.06 &  7.15 &  0.57  \\
CES\,1543$+$0201 &  $-$2.65 & $-$2.51    & 6.23 &  0.49    &  4.57 & $-$0.64 &    -- &   --   \\
CES\,1552$+$0517 &  $-$2.60 & $-$2.49    & 6.07 &  0.28    &  5.76 &  0.50   &    -- &   --   \\
CES\,1732$+$2344 &  $-$2.57 & $-$2.67    & 6.37 &  0.55    &    -- &    --   &    -- &   --   \\
CES\,1804$+$0346 &  $-$2.48 & $-$2.43    &   -- &    --    &  6.71 &  1.33   &  7.07 &  0.82  \\
CES\,1942$-$6103 &  $-$3.34 & $-$3.14    & 4.90 & $-$0.15  &  5.47 &  0.95   &    -- &   --   \\
CES\,2019$-$6130 &  $-$2.97 & $-$2.92    & 5.65 &  0.23    &    -- &    --   &    -- &   --   \\
CES\,2103$-$6505 &  $-$3.58 & $-$3.20    &   -- &    --    &  4.35 &  0.07   &    -- &   --   \\
CES\,2231$-$3238 &  $-$2.77 & $-$2.70    & 6.08 &  0.46    &  4.94 & $-$0.15 &    -- &   --   \\
CES\,2232$-$4138 &  $-$2.58 & $-$2.47    & 6.23 &  0.42    &  4.60 & $-$0.68 &    -- &   --   \\
CES\,2250$-$4057 &  $-$2.14 & $-$1.88    & 6.43 &  0.18    &  5.89 &  0.17   &    -- &   --   \\
CES\,2254$-$4209 &  $-$2.88 & $-$2.64    & 5.39 & $-$0.12  &  5.27 &  0.29   &    -- &   --   \\
CES\,2330$-$5626 &  $-$3.10 & $-$2.97    & 5.87 &  0.58    &  4.50 & $-$0.26 &    -- &   --   \\
CES\,2334$-$2642 &  $-$3.48 & $-$3.39    & 5.03 &  0.12    &    -- &    --   &    -- &   --   \\
\hline
\end{tabular}
\label{tab:abundances}
\end{table*}

\newpage

\section{Atmospheric parameters}

\begin{table*}[h!]
    \centering
    \caption{Stellar parameters from \cite{2022A&A...665A..10L} for the stars in our sample.}
\begin{tabular}{ccccc}
\hline\hline
Star &  $T_\mathrm{eff}$ & $\mathrm{log}\,g$ & [Fe/H] & vt \\ 
 & K & dex & dex & km$\,$s$^{-1}$ \\\hline
CES\,0031$-$1647 &  4960 & 1.83 & $-$2.49 & 1.91  \\
CES\,0045$-$0932 &  5023 & 2.29 & $-$2.95 & 1.76  \\
CES\,0048$-$1041 &  4856 & 1.68 & $-$2.48 & 1.93  \\
CES\,0055$-$3345 &  5056 & 2.45 & $-$2.36 & 1.66  \\
CES\,0059$-$4524 &  5129 & 2.72 & $-$2.39 & 1.56  \\
CES\,0102$-$6143 &  5083 & 2.37 & $-$2.86 & 1.75  \\
CES\,0107$-$6125 &  5286 & 2.97 & $-$2.59 & 1.54  \\
CES\,0109$-$0443 &  5206 & 2.74 & $-$3.23 & 1.69  \\
CES\,0215$-$2554 &  5077 & 2.00 & $-$2.73 & 1.91  \\
CES\,0221$-$2130 &  4908 & 1.84 & $-$1.99 & 1.84  \\
CES\,0242$-$0754 &  4713 & 1.36 & $-$2.90 & 2.03  \\
CES\,0301$+$0616 &  5224 & 3.01 & $-$2.93 & 1.51  \\
CES\,0338$-$2402 &  5244 & 2.78 & $-$2.81 & 1.62  \\
CES\,0413$+$0636 &  4512 & 1.10 & $-$2.24 & 2.01  \\
CES\,0419$-$3651 &  5092 & 2.29 & $-$2.81 & 1.78  \\
CES\,0422$-$3715 &  5104 & 2.46 & $-$2.45 & 1.68  \\
CES\,0424$-$1501 &  4646 & 1.74 & $-$1.79 & 1.74  \\
CES\,0430$-$1334 &  5636 & 3.07 & $-$2.09 & 1.63  \\
CES\,0444$-$1228 &  4575 & 1.40 & $-$2.54 & 1.92  \\
CES\,0518$-$3817 &  5291 & 3.06 & $-$2.49 & 1.49  \\
CES\,0527$-$2052 &  4772 & 1.81 & $-$2.75 & 1.84  \\
CES\,0547$-$1739 &  4345 & 0.90 & $-$2.05 & 2.01  \\
CES\,0747$-$0405 &  4111 & 0.54 & $-$2.25 & 2.08  \\
CES\,0900$-$6222 &  4329 & 0.94 & $-$2.11 & 1.98  \\
CES\,0908$-$6607 &  4489 & 0.90 & $-$2.62 & 2.12  \\
CES\,0919$-$6958 &  4430 & 0.70 & $-$2.46 & 2.17  \\
CES\,1116$-$7250 &  4106 & 0.48 & $-$2.74 & 2.14  \\
CES\,1221$-$0328 &  5145 & 2.76 & $-$2.96 & 1.60  \\
CES\,1222$+$1136 &  4832 & 1.72 & $-$2.91 & 1.93  \\
CES\,1226$+$0518 &  5341 & 2.84 & $-$2.38 & 1.60  \\
CES\,1228$+$1220 &  5089 & 2.04 & $-$2.32 & 1.87  \\
CES\,1237$+$1922 &  4960 & 1.86 & $-$3.19 & 1.95  \\
CES\,1245$-$2425 &  5023 & 2.35 & $-$2.85 & 1.72  \\
CES\,1322$-$1355 &  4960 & 1.81 & $-$2.93 & 1.96  \\
CES\,1402$+$0941 &  4682 & 1.35 & $-$2.79 & 2.01  \\
CES\,1405$-$1451 &  4642 & 1.58 & $-$1.87 & 1.81  \\
CES\,1413$-$7609 &  4782 & 1.72 & $-$2.52 & 1.87  \\
CES\,1427$-$2214 &  4913 & 1.99 & $-$3.05 & 1.85  \\
CES\,1436$-$2906 &  5280 & 3.15 & $-$2.15 & 1.42  \\
CES\,1543$+$0201 &  5157 & 2.77 & $-$2.65 & 1.57  \\
CES\,1552$+$0517 &  5013 & 2.30 & $-$2.60 & 1.72  \\
CES\,1732$+$2344 &  5370 & 2.82 & $-$2.57 & 1.65  \\
CES\,1804$+$0346 &  4390 & 0.80 & $-$2.48 & 2.12  \\
CES\,1942$-$6103 &  4748 & 1.53 & $-$3.34 & 2.01  \\
CES\,2019$-$6130 &  4590 & 1.13 & $-$2.97 & 2.09  \\
CES\,2103$-$6505 &  4916 & 2.05 & $-$3.58 & 1.85  \\
CES\,2231$-$3238 &  5222 & 2.67 & $-$2.77 & 1.67  \\
CES\,2232$-$4138 &  5194 & 2.76 & $-$2.58 & 1.59  \\
CES\,2250$-$4057 &  5634 & 2.51 & $-$2.14 & 1.88  \\
CES\,2254$-$4209 &  4805 & 1.98 & $-$2.88 & 1.79  \\
CES\,2330$-$5626 &  5028 & 2.31 & $-$3.10 & 1.75  \\
CES\,2334$-$2642 &  4640 & 1.42 & $-$3.48 & 2.02  \\
\hline
\end{tabular}
\label{parameters}
\end{table*}

\newpage
\section{Slopes}

\begin{table*}[h!]
    \centering
    \caption{Slopes for the s-elements Sr, Y, and Zr with respect to the [C/N] ratios.}
\begin{tabular}{cccc}
\hline\hline
Element & slope$\pm$SE  & slope$\pm$SE & slope$\pm$SE \\
& mixed & unmixed & mixed and unmixed\\\hline

Sr & 0.191$\pm$0.276 & -0.020$\pm$0.253  & -0.075$\pm$0.085 \\
Y  &  0.242$\pm$0.286 & 0.018$\pm$0.158 & -0.0007$\pm$0.07282 \\
Zr & 0.148$\pm$0.375 & 0.058$\pm$0.155 & 0.008$\pm$0.084 \\

\hline
\end{tabular}
\tablefoot{SE represents the standard error on the angular coefficients.}
\label{slopes}
\end{table*}

\section{Supplementary figures}

\begin{figure*}[h!]
\centering
   \includegraphics[width=18cm]{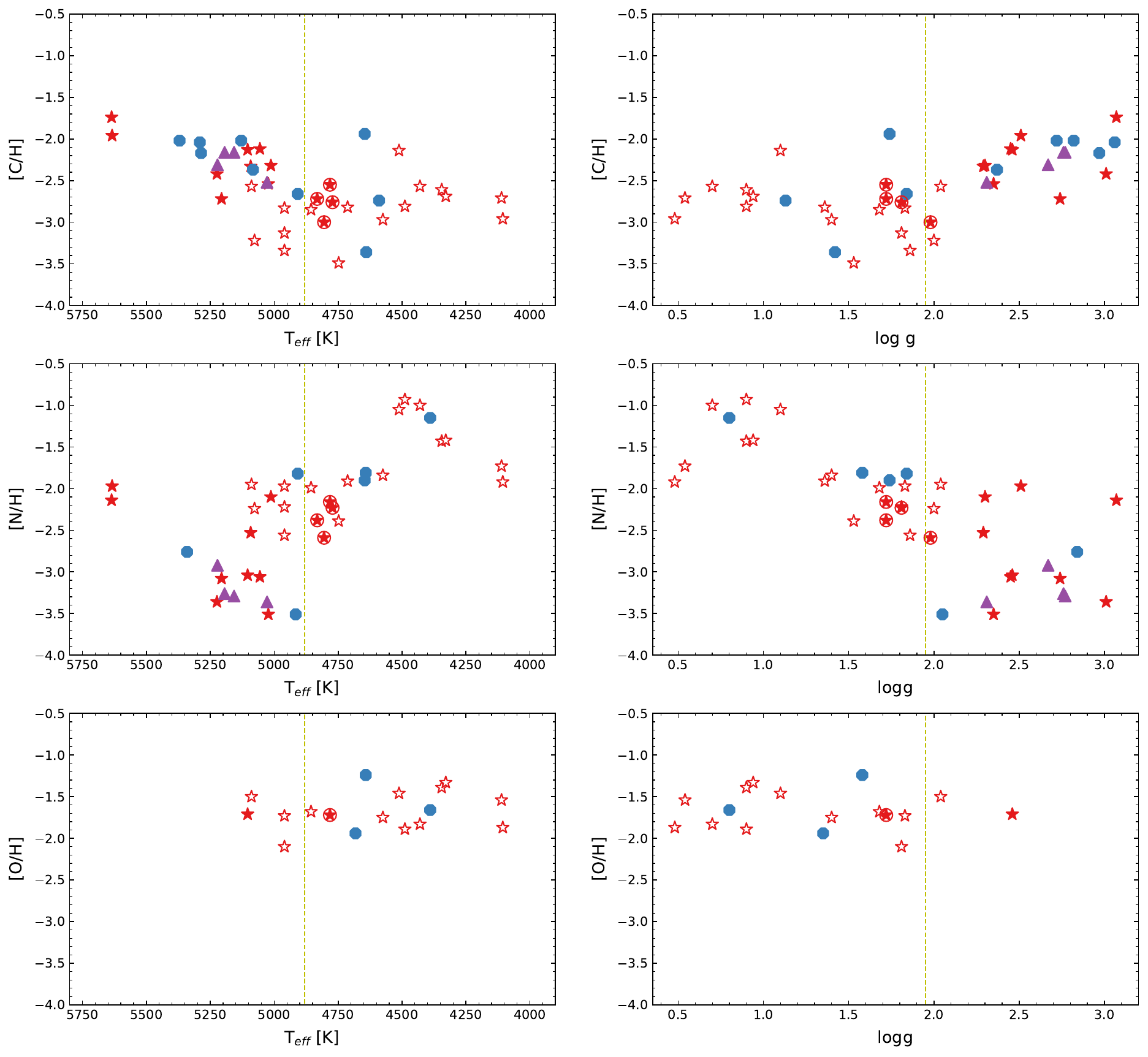}
  \caption{ [C/H] as a function of the effective temperature (on the left) and the surface gravity (on the right) for all the stars in our sample (top). A clear separation can be seen between the mixed and unmixed stars, with the exception of the four encircled stars. [N/H] with respect to $T_\mathrm{eff}$ and $\mathrm{log}\,g$ are shown in the middle panels. The lower panel shows [O/H] as a function of $T_\mathrm{eff}$ and $\mathrm{log}\,g$. The yellow dashed line indicates the $T_\mathrm{eff}$ and $\mathrm{log}\,g$ at the RGB bump. Symbols are the same as in Fig.\ref{fig:nfe_cfe}. Using H instead of Fe makes our unmixed stars are spreader, however it does not vanish the separation between mixed and unmixed.}
  \label{fig:ch_nh_teff_logg}
\end{figure*}

\twocolumn 

\begin{figure}
  \resizebox{8cm}{!}{\includegraphics{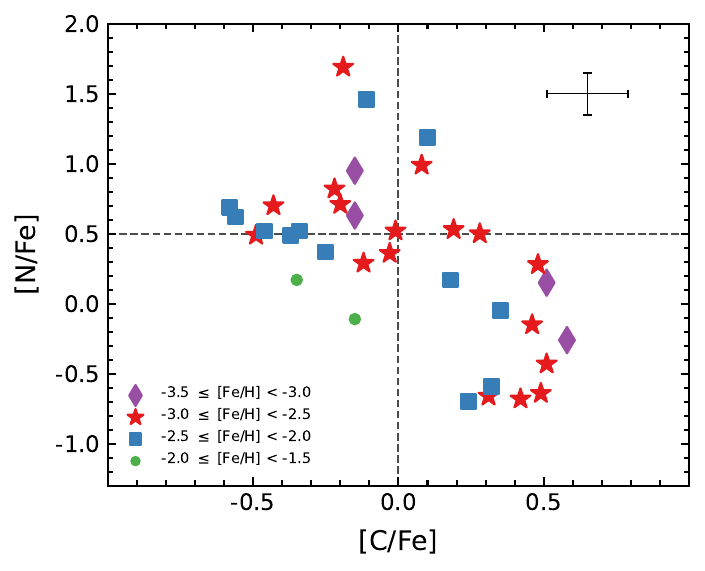}}
  \caption{[N/Fe] versus [C/Fe] for our sample stars divided in metallicity bins (see the legend in the bottom left corner of the figure). We could not find any trends that could indicate a dependence of stellar mixing on metallicity. However, the stars we could not classify following the adopted limits are the most metal-rich ones of our sample. A representative error bar is plotted in the upper
right corner of the figure.}
  \label{fig:ncfe_wmet}
\end{figure}

\begin{figure*}
  \resizebox{\hsize}{!}{\includegraphics{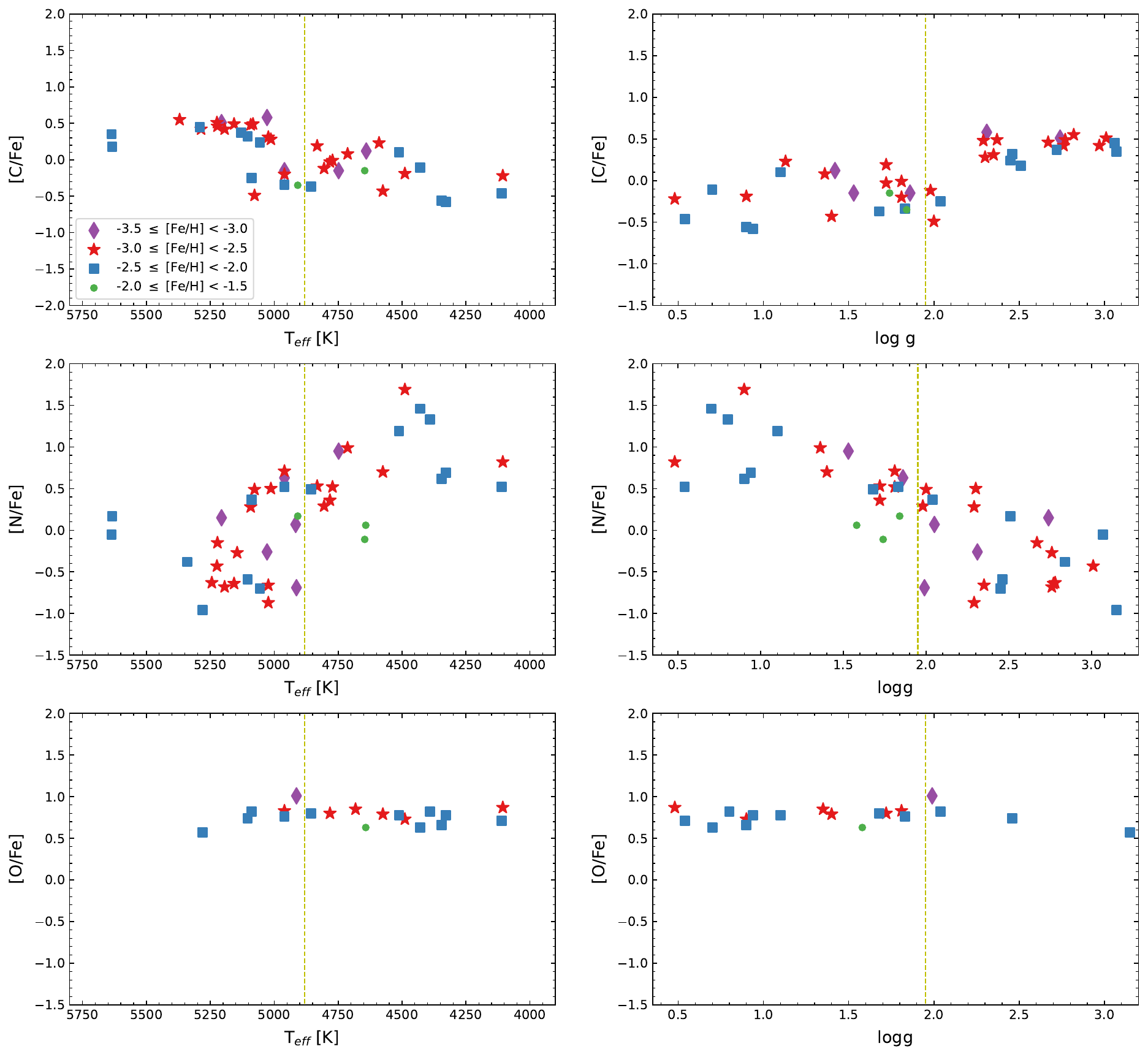}}
  \caption{ [C/Fe] as a function of the effective temperature (on the left) and the surface gravity (on the right) for all the stars in our sample (top). [N/Fe] with respect to $T_\mathrm{eff}$ and $\mathrm{log}\,g$ are shown in the middle panels. The lower panel shows [O/H] as a function of $T_\mathrm{eff}$ and $\mathrm{log}\,g$. The yellow dashed line indicates the $T_\mathrm{eff}$ and $\mathrm{log}\,g$ at the RGB bump. As in Fig.~\ref{fig:ncfe_wmet}, by plotting for different bins of metallicity, we could not find trends for the mixing and unmixed stars. }
  \label{fig:cfe_nfe_logg_teff_wmet}
\end{figure*}

\end{appendix}
\end{document}